\documentclass[aps, prl, twocolumn, longbibliography]{revtex4-2}
\usepackage{graphicx}
\usepackage{amsmath}
\usepackage{amssymb}
\usepackage{color}
\usepackage{appendix}
\usepackage{braket}
\usepackage{bbm}
\usepackage{subfigure}
\usepackage{mathrsfs}

\usepackage[colorlinks,urlcolor=blue,citecolor=blue,linkcolor=blue]{hyperref}
\usepackage{lipsum}
\usepackage{diagbox}
\usepackage{multirow}

\begin{document}
\title{Anatomy of Non-Hermitian Dynamical Quantum Phase Transitions}

\author{Yongxu Fu}
\email{yongxufu@zjnu.edu.cn}
\affiliation{Department of Physics, Zhejiang Normal University, Jinhua 321004, China}

\author{Gao Xianlong}
\email{gaoxl@zjnu.edu.cn}
\affiliation{Department of Physics, Zhejiang Normal University, Jinhua 321004, China}

\begin{abstract}
We establish a unified framework for dynamical quantum phase transitions (DQPTs) in non-Hermitian systems that encompasses both biorthogonal and self-norm non-biorthogonal formulations for pure and mixed states under quantum quench protocols. Our framework provides explicit expressions for the Loschmidt amplitude, Loschmidt echo, and rate function, revealing a universal geometric signature of DQPTs in the two-band model: orthogonality of two related vectors in two-dimensional real space. Strikingly, we demonstrate that non-biorthogonal quenches from non-Hermitian to Hermitian Hamiltonians under chiral symmetry exhibit emergent topological characteristics of DQPTs, unveiling their fundamental features beyond conventional Hermitian regimes. This work establishes fundamental geometric and topological principles governing quantum criticality in open systems, with implications for quantum sensing and many-body physics in dissipative environments.
\end{abstract}

\maketitle

\emph{Introduction.---}
The classification and understanding of different phases of matter, along with the mechanisms driving their transitions, have long been central pursuits in condensed matter physics \cite{Altland_Simons_2010, mahan2000book, negele2018book}. Importantly, equilibrium thermal and quantum phase transitions are now described by mature and comprehensive theoretical formalisms \cite{sondhi1997con, Sachdev_2011}. However, nonequilibrium systems, which are more ubiquitous in nature, pose greater challenges for the study of phase transitions. This is largely due to the inherent openness and complexity of such systems, which complicate their theoretical description and obscure universal critical behavior \cite{fischer2006sweeping, fischer2007vortex, noneq1, noneq2, noneq3, noneq4}. Over the past decade, a theoretical protocol based on the return probability of nonequilibrium quantum states---particularly formulated as the Loschmidt echos in quantum quench scenarios---has revealed a novel class of phase transitions in nonequilibrium systems, known as dynamical quantum phase transitions (DQPTs) \cite{heyl2013,heyl2014,heyl2015,review2016,review2018}. These transitions may exhibit features reminiscent of equilibrium or topological phase transitions \cite{canovi2014first,vajna2015,heyl2016,lang2018}. A surge of research on DQPTs has since emerged, encompassing both theoretical investigations \cite{vajna2014disentangling,palmai2015,divakaran2016,sharma2016,sedlmayr2018,wang2018connection, pastori2020,sadrazdeh2021,corps2023,van2023,jafari2024,maslowski2023,xiao2024dyn} and experimental observations \cite{jurcevic2017,garttner2017,zhang2017obser,flaschner2018,tian2019obser,guo2019obser,wang2019simu,nie2020,chen2020,xu2020measuring,hamazaki2021,sun2021uncover}. Remarkably, DQPTs in mixed-state (finite-temperature) systems also uncover underlying topological features, extending the concept of DQPTs beyond pure-state dynamics \cite{abeling2016mixed,heyl2017mixed,bhattacharya2017mixed,lang2018mixed,mera2018mixed,hou2020mixed,hou2022mixed}. These developments are closely relevant to the Uhlmann phase \cite{uhlmann1986,uhlmann1989,uhlmann1991,hubner1992,hubner1993,ajoqvist2000,viyuela2014}, which serves as a mixed-state generalization of the Berry phase \cite{simon1983,berryphase1984,aharonov1987} or Pancharatnam phase \cite{pancharatnam1956, samuel1988, tong2004kin, cohen2019} defined for pure states.  

In recent years, the foundational framework of quantum and condensed matter physics, particularly the Hermiticity of the Hamiltonian, has been profoundly extended to involve non-Hermitian systems \cite{ashida2020,bergholtzrev2021,gong2018,kawabataprx,yao2018,yao201802,yokomizo2019, slager2020,yang2020,origin2020, wang2024amoeba}, thereby reshaping the conventional Bloch-based formalism. The rapid theoretical development of non-Hermitian quantum systems has inevitably impacted the study of DQPTs, prompting the exploration of their generalizations in non-Hermitian settings \cite{sefdlmayr2018fate,lang2018open,zhou2018dyn,zhou2021non,tzeng2021,tang2022dyn,naji2022dyn,mondal2022dyn,mondal2023dyn,jing2024dyn, agarwal2024detecting, agarwal2023recognizing}. Given that the diagonalization of a non-Hermitian (non-defective) Hamiltonian yields biorthogonal eigenvectors, it is natural that DQPTs in non-Hermitian systems---particularly those formulated via the Loschmidt amplitudes and echos---should be evaluated within the biorthogonal framework \cite{mondal2022dyn,mondal2023dyn,jing2024dyn}. However, Loschmidt echos defined within the non-biorthogonal (self-norm) formalism often exhibit a more intuitive physical interpretation, as they directly represent the return probability after a quantum quench \cite{zhou2018dyn,zhou2021non}, similar to the Hermitian cases. 

Motivated by these considerations, this work presents a comprehensive anatomy of the non-Hermitian DQPTs, formulated in both biorthogonal and non-biorthogonal, pure-state and mixed-state formalisms, grounded in fundamental principles from condensed matter physics, statistical mechanics, and non-Hermitian quantum theory. We develop a unified and systematic formulation of Loschmidt amplitudes and echos for generic mixed states, which naturally recovers the pure-state expressions when the system is initialized in its ground state. Bases on biorthogonal and non-biorthogonal frameworks, we explicitly identify the conditions under which DQPTs occur, revealing their inherent geometric character in the two-band model. Surprisingly, the topological characteristics of the DQPTs emerge in a specific non-biorthogonal quantum quench scenario.

\emph{Mixed-state Loschmidt amplitudes in non-Hermitian systems.---}
A fundamental approach to characterizing the finite-temperature DQPTs in Hermitian systems employs the mixed-state Loschmidt amplitude upon the quantum quench protocol starting from $t=0$  \cite{heyl2017mixed,bhattacharya2017mixed,lang2018mixed,mera2018mixed,hou2020mixed,hou2022mixed}
\begin{align}
    \label{eqfinitetlsam}
    \mathcal{G}(t)=\operatorname{Tr}\left[\rho(0)U(t)\right],
\end{align}
where $\rho(0)$ is the initial thermal density matrix corresponding to the prequench Hamiltonian $\hat{H}_{0}$, and $U(t)=\exp (-i\hat{H}_{1}t)$ is the unitary time-evolution operator governed by the postquench Hamiltonian $\hat{H}_{1}$. In general, $U(t)$ does not correspond to the parallel transport on a principal fiber bundle with the unitary structure group $\mathbb{U}(\mathcal{N})$, and thus lacks gauge invariance. Here, $\mathcal{N}$ denotes the dimension of the Hilbert space at each point of the base manifold, as reviewed in the Supplementary Materials \cite{supp} for details (see also references \cite{du2022pt, yin2018geometric, yang2024complex, cao2024topo,lu2025dyn, budich2015topo, cao2023statistical} therein). A $\mathbb{U}(1)$-invariant geometric phase, which has been proposed to signal the occurrence of DQPTs \cite{bhattacharya2017mixed,heyl2017mixed}, can be defined in Hermitian systems but is difficult to rigorously generalize to the non-Hermitian case. Although certain numerical studies have demonstrated the geometric phase signatures in non-Hermitian DQPTs \cite{zhou2018dyn, zhou2021non, mondal2022dyn, mondal2023dyn, jing2024dyn}, the present work focuses on elucidating the fundamental theoretical connection between Loschmidt amplitudes and DQPTs for non-Hermitian systems \cite{supp}.

We demonstrate that the definition of the generic Loschmidt amplitude in Eq. (\ref{eqfinitetlsam}) can be feasibly generalized to non-interacting fermionic non-Hermitian systems, where both $\hat{H}_{0}$ and $\hat{H}_{1}$ are generic non-Hermitian Hamiltonians \footnote{Although an alternative definition of the mixed-state Loschmidt amplitude can be formulated based on the parallel-transport trajectory of the time-evolved density matrix \cite{tong2004kin, lang2018open}, its extension to non-Hermitian Hamiltonians remains challenging, as most quantities in non-Hermitian systems---including the participation probabilities of individual states in mixed states---are complex.}. This generalization is physically transparent, as it directly evaluates the expectation value of the postquench time-evolution operator with respect to the initial prequench state, up to the normalization factor $\operatorname{Tr}\left[\rho(0)\right]$. However, from a mathematical perspective, the density matrix of a non-Hermitian Hamiltonian is formulated in the biorthogonal basis as
\begin{align}
    \varrho=\sum_{\alpha}\ket{\Psi_{\alpha}}p_{\alpha}\langle\!\langle\Psi_{\alpha}| \label{eqbdm},
\end{align}
where $\ket{\Psi_{\alpha}}, |\Psi_{\alpha}\rangle\!\rangle$ denote the many-body right and left eigenstates, respectively, each constructed from arbitrary occupations of single-particle states. The coefficients $p_{\alpha}$ represent the participation probabilities associated with these eigenstates. It is important to note that $p_{\alpha}$ can take complex values, which implies that the non-negative trace-class property of the density matrices cannot be guaranteed in non-Hermitian systems. Although complex participation probabilities lack direct physical significance, the inherent ambiguity in the physical interpretation of non-Hermitian Hamiltonians compels us to focus on derived quantities obtained through further analysis to quantify their physical relevance. In this paper, we focus on one-dimensional systems; however, the general formalism remains applicable to higher-dimensional settings.

Incorporating the biorthogonal formulation, the Loschmidt amplitude in Eq. (\ref{eqfinitetlsam}) can be factorized into a product over specific momentum sectors under periodic boundary conditions, 
\begin{align}
    \label{eqlsaexp}
    \mathcal{G}(t)=\prod_{k \in occ}\mathcal{G}_{k}(t), \quad \mathcal{G}_{k}(t)\equiv \mathrm{Tr}\left[\varrho_{k}(0)U_{k}(t)\right],
\end{align}
given a specified fermionic occupation of the Brillouin zone (BZ) \footnote{If no specific occupation of the BZ is predetermined, a summation over all possible momentum configurations is required, along with appropriate symmetrization or anti-symmetrization depending on whether the particles are bosons or fermions. This results in density matrices and Loschmidt amplitudes that are cumbersome and not easily tractable. In this work, we focus on specified fermionic occupations (with non-repetitive $k$) of the BZ in both the biorthogonal and non-biorthogonal formulations of DQPTs.}, where $U_{k}(t)=\exp [-iH_{1}(k)t]$ and $H_{1}(k)$ is the single-particle postquench Hamiltonian in momentum space. The initial density matrix $\varrho(0)$ decomposes into several single-particle components 
\begin{align}
    \label{eqfactorden}
    \varrho_{k}(0)=\sum_{n}\ket{u_{n}^{0}(k)}p_{k}^{n}\langle\!\langle u_{n}^{0}(k)|,
\end{align}
where $\ket{u_{n}^{0}(k)}$ and $|u_{n}^{0}(k)\rangle\!\rangle$ are the $n$th-band right and left eigenstates of the single-particle prequench Hamiltonian $H_{0}(k)$, respectively. The many-body probabilities $p_{\alpha}$ are expressed as a product over single-particle contributions, $p_{\alpha}=\prod_{(k,n)} p_{k}^{n}$, corresponding to a given occupation configuration (see Supplementary Materials \cite{supp} for detail calculations). The Loschmidt amplitude in Eq. (\ref{eqlsaexp}) applies to initial mixed states with arbitrary band occupation configurations that are compatible with the specified momentum occupation. In particular, when $p_{k}^{n}$ is nonzero for all bands, the quasi-momentum $k$ spans half or the entire BZ, corresponding to a half or full filled system, respectively. Moreover, the corresponding Loschmidt echo is defined as the squared norm of the Loschmidt amplitude, $\mathcal{L}(t)=\left|\mathcal{G}(t)\right|^{2}$, up to an overall normalization factor that does not affect the occurrence of DQPTs \footnote{Throughout this paper, we omit such normalization factors in the expressions of the Loschmidt amplitude and echo for simplicity, as they are irrelevant to the emergence of DQPTs.}.

Mixed-state Loschmidt amplitude in Eq. (\ref{eqlsaexp}) reduces to the pure-state case when only a single participation probability 
$p_{\alpha}$ is nonzero---particularly, when it corresponds to the biorthogonal ground states $\ket{\Psi_{0}}$ and $|\Psi_{0}\rangle\!\rangle$ in the zero-temperature limit. For two-band ($n=\pm$) systems with the lower band occupied, the Loschmidt amplitude reduces to 
\begin{align}
    \label{eqpurelsa}
    \mathcal{G}_{ZT}(t)=\prod_{k\in occ}\langle\!\langle u_{-}^{0}(k)\ket{u_{-}^{0}(k,t)},
\end{align}
where $\ket{u_{-}^{0}(k,t)}\equiv e^{-iH_{1}(k)t}\ket{u_{-}^{0}(k)}$. The corresponding Loschmidt echo, defined as the squared norm of $\mathcal{G}_{ZT}(t)$, is given by
\begin{align}
    \label{eqlse}
    \mathcal{L}_{ZT}(t)=\prod_{k\in occ}\langle\!\langle u_{-}^{0}(k)\ket{u_{-}^{0}(k,t)}\langle\!\langle u_{-}^{0}(k,t)\ket{u_{-}^{0}(k)},
\end{align}
which is consistent with the expression in Ref. \cite{jing2024dyn}, up to a normalization factor $\langle\!\langle u_{-}^{0}(k,t)\ket{u_{-}^{0}(k,t)}^{-1}$. Here, $|u_{-}^{0}(k,t)\rangle\!\rangle$ denotes the associated state of $\ket{u_{-}^{0}(k,t)}$ (see Supplementary Materials \cite{supp} for details).

\emph{Biorthogonal DQPTs.---}
In the context of biorthogonal DQPTs, the corresponding rate function, which plays a role analogous to the free energy density in equilibrium phase transitions, is given by
\begin{align}
    \label{eqlsr}
    \lambda(t)=-\lim_{N\rightarrow \infty}\frac{1}{N}\ln \mathcal{L}(t),
\end{align}
where $N$ denotes the number of degrees of freedom. Rate function nonanalyticity signals DQPTs via Fisher zeros of the Loschmidt amplitude \cite{heyl2013, review2016, review2018}. 

We focus on a generic gapped two-band lattice model described by $\mathcal{H}(k)=\vec{h}(k) \cdot \vec{\sigma}$, where $\vec{\sigma}=(\sigma_{x}, \sigma_{y}, \sigma_{z})$ is the vector of three Pauli matrices. The prequench and postquench Hamiltonians, $\mathcal{H}_{0}(k)$ and $\mathcal{H}_{1}(k)$, correspond to different parameterizations of $\vec{h}_{0}(k)$ and $\vec{h}_{1}(k)$, respectively. The rate function can be expressed as \footnote{In two-band models, the degrees of freedom are twice the number of unit cells, so the summation-to-integration coefficient is $1/4\pi$ in principle, which effectively serves as a part of the overall normalization factor and has no essential influence on the occurrence of DQPTs. In practical calculations, the overall factor can be flexibly chosen as needed to confine the rate function within the range $[0,1]$.}
\begin{align}
    \label{eqratefync}
    \lambda(t)=- \frac{1}{4\pi}\int dk \ln (\left|\mathcal{G}_{k}(t)\right|^{2}),
\end{align}
in the thermodynamic limit, where the integration range depends on the specific occupation configuration. We begin with the initial generalized Gibbs state at temperature $T=1/\beta$,
\begin{align}
    \label{eqgibbs}
    \varrho_{k}(0)=\frac{e^{-\beta \mathcal{H}_{0}(k)}}{\mathrm{Tr} \left[e^{-\beta \mathcal{H}_{0}(k)} \right]}=\frac{1}{2}\left[\mathbbm{1}_{2 \times 2}-n_{k}\hat{h}_{0}(k)\cdot \vec{\sigma}\right],
\end{align}
where $h_{0}(k)=\left[h_{0,x}^{2}(k)+h_{0,y}^{2}(k)+h_{0,z}^{2}(k)\right]^{1/2}$, $\hat{h}_{0}(k)=\vec{h}_{0}(k)/h_{0}(k)$, and $n_{k}=\tanh\left[\beta h_{0}(k)\right]$ is generally complex value. This mixed state formally has complex participation probabilities $p_{k}^{\pm}=\frac{1}{2}\left(1\mp n_{k} \right)$,
associated with the two bands $\epsilon_{k\pm}^{0}=\pm h_{0}(k)$. The Loschmidt amplitude factor for each momentum mode $k$ is given by 
\begin{align}
    \label{eqlsatwoband}
    \mathcal{G}_{k}(t)=\cos\left[h_{1}(k)t\right]+i\sin\left[h_{1}(k)t\right]n_{k}\hat{h}_{0} \cdot \hat{h}_{1}
\end{align}
where $\hat{h}_{1}(k)$ is defined analogously to $\hat{h}_{0}(k)$ as the normalized vector $\vec{h}_{1}(k)/h_{1}(k)$. In the zero-temperature limit $\beta \rightarrow \infty$ (corresponding to the pure ground-state case), we have $n_{k} = \tanh[\beta h_{0}(k)] \rightarrow 1$, and the Loschmidt amplitude factor simplifies to $\mathcal{G}^{ZT}_{k}(t)=\cos\left[h_{1}(k)t\right]+i\sin\left[h_{1}(k)t\right]\hat{h}_{0} \cdot \hat{h}_{1}$. In the infinite-temperature limit $\beta \rightarrow 0$, the Loschmidt amplitude reduces to $\mathcal{G}^{IT}_{k}(t)=\cos\left[h_{1}(k)t\right]$, which coincides with the case studied in Ref.~\cite{zhou2021non}.

Analogous to the Lee-Yang zeros of thermodynamic functions \cite{yang1952, lee1952}, the time variable $t$ in $\mathcal{G}(t)$ can be extended into the complex plane via the analytic continuation $t \mapsto z \in \mathbb{C}$. The Fisher zeros of the analytically continued $\mathcal{G}_{k}(z)$ are then given by (for integer $n$)
\begin{align}
    z_{n}=\frac{i\pi(2n+1)}{2 h_{1}(k)}-\frac{1}{h_{1}(k)}\tanh^{-1}\left(n_{k}\hat{h}_{0} \cdot \hat{h}_{1}\right).
\end{align}
Introducing the notation $\mathcal{Q}_{k}=\tanh^{-1}\left(n_{k}\hat{h}_{0} \cdot \hat{h}_{1}\right)$ and $a=\pi(n+1/2)$, and decomposing 
$h_{1}(k)$ and $\mathcal{Q}_{k}$ into their real and imaginary parts as $h_{1r}=\mathrm{Re}\left[h_{1}(k)\right],h_{1i}=\mathrm{Im}\left[h_{1}(k)\right],\mathcal{Q}_{kr}=\mathrm{Re}\left[\mathcal{Q}_{k}\right], \mathcal{Q}_{ki}=\mathrm{Im}\left[\mathcal{Q}_{k}\right]$, the real and imaginary parts of the Fisher zeros can be identified.
The physical critical points of DQPTs are determined by the Fisher zeros lying on the imaginary axis, which simultaneously identify the critical momenta 
$k_c$ through the condition $a h_{1i}-h_{1r}\mathcal{Q}_{kr}-h_{1i}\mathcal{Q}_{ki}=0$. This condition implies that the two vectors $\vec{v}_{s}=(\mathcal{Q}_{kr},\mathcal{Q}_{ki}-a)$ and $\vec{v}_{h}=(h_{1r},h_{1i})$ in the two-dimensional (2D) plane are orthogonal, i.e., 
\begin{align}
    \label{eq-condition}
    \vec{v}_{s}\cdot \vec{v}_{h}=0.
\end{align}
Hence, as long as the polar angle between this two vectors $\phi = \phi_{s}- \phi_{h}$ in the 2D plane sweeps through $m \pi/2$ with odd $m$, the critical point $k_{c}$ of the DQPTs will occur, where $\phi_{s}$ and $\phi_{h}$ denote the polar angles of $\vec{v}_{s}$ and $\vec{v}_{h}$, respectively (with the $k$-dependence omitted for brevity). Consequently, the DQPT critical times $t_{c}$ correspond to the sequence of $z_{n}$ evaluated at the critical momenta $k_{c}$, where the condition $\vec{v}_{s}\cdot \vec{v}_{h}=0$ is satisfied. 

From a geometric perspective, the winding number associated with a generic $k$-dependent 2D vector is computed by integrating its polar angle $\varphi$ over the BZ, 
\begin{align}
    \label{eqwinding}
    W=\frac{1}{2\pi}\oint_{\mathrm{BZ}}d\varphi.
\end{align}
Define $\Delta W = W_{s} - W_{h}$ as the difference between the winding numbers associated with $\vec{v}_{s}$ and $\vec{v}_{h}$, respectively, and if 
\begin{align}
    \label{eqwindiff}
    \left|\Delta W \right| \geq \frac{1}{2},
\end{align}
then the phase difference $\phi$ must traverse at least one multiple of $m\pi/2$ (with odd $m$) across the BZ, regardless of the initial offset between $\phi_{s}$ and $\phi_{h}$, thereby ensuring the occurrence of a DQPT. However, the occurrence of DQPTs does not guarantee that Eq. (\ref{eqwindiff}) holds, due to the arbitrary initial offset of $\phi$. In this context, the conditions for the occurrence of biorthogonal DQPTs reflect a geometric rather than a topological feature, owing to the generic non-topological (quantized) nature of $\Delta W$. 

The topological character manifests in the Hermitian limit with certain symmetries, where the condition in Eq. (\ref{eq-condition}) reduces to $\hat{h}_{0} \cdot \hat{h}_{1}=0$ in zero or finite temperature, and becomes directly connected to underlying topological invariants characterizing the (equilibrium) topological phases before and after the quench, such as the winding number, $\mathbb{Z}_{2}$ invariant, and Chern number \cite{vajna2015, heyl2016, bhattacharya2017mixed, heyl2017mixed}. In the infinite-temperature limit, DQPTs necessarily occur in Hermitian systems due to the real-value nature of $h_{1}(k)$, whereas in intrinsically non-Hermitian systems, their occurrence becomes conditional. Interestingly, while the DQPTs in Hermitian systems do not depend upon the initial-state (zero or finite) temperature \cite{bhattacharya2017mixed, heyl2017mixed}, temperature plays a nontrivial role in non-Hermitian DQPTs due to the complex-valued nature of the Hamiltonians. In other words, complex Hamiltonians, which leads to the increased complexity of the generic condition for DQPTs in Eq. (\ref{eq-condition}), enhance the likelihood of the DQPTs occurring not only at zero temperature but also at finite temperatures, compared with the Hermitian cases.

\emph{Non-biorthogonal DQPTs.---}
The biorthogonal formulation provides a natural mathematical generalization from Hermitian to non-Hermitian scenarios. Nonetheless, biorthogonality is not always essential, as its necessity may depend on specific physical considerations. Particularly in the pure-state case, the non-biorthogonal Loschmidt amplitude 
\begin{align}
    \mathscr{G}_{ZT}(t)=\prod_{(n,k)\in occ}\braket{u_{n}^{0}(k)|u_{n}^{0}(k,t)},
\end{align}
and the associated Loschmidt echo $\mathscr{L}_{ZT}(t)=\left|\mathscr{G}_{ZT}(t)\right|^{2}$ directly quantify the overlap between the time-evolved postquench state and the initial eigenstate, characterizing the return probability following a quantum quench. Additionally, the non-biorthogonal observables, normalized by the norm of the associated time-evolved state, align with the no-jump limit of the approximate Lindblad master equation \cite{sim2023quantum, sim2025onservables}. Formally, in the non-biorthogonal formulation of DQPTs, the relevant expressions differ from the biorthogonal case only by replacing the terms involving left eigenstates with the Hermitian conjugates of the corresponding right eigenstates. Thus, the mixed-state non-biorthogonal Loschmidt amplitude is given by
\begin{align}
    \mathscr{G}(t)=\prod_{k \in occ}\mathscr{G}_{k}(t), \quad \mathscr{G}_{k}(t)\equiv \mathrm{Tr}\left[\rho_{k}(0)U_{k}(t)\right],
\end{align}
with a specified fermionic occupation of the BZ, where the single-particle components of the non-biorthogonal initial density matrix $\rho(0)$ is
\begin{align}
    \rho_{k}(0)=\sum_{n}p_{k}^{n}\ket{u_{n}^{0}(k)}\bra{u_{n}^{0}(k)}.
\end{align}

We again take the two-band model $\mathcal{H}(k)$ as a representative example for illustration. However, in the non-Hermitian context, a non-biorthogonal mixed state such as the Gibbs state---with analytically tractable participation probabilities $p_{k}^{\pm}$---is generally not applicable. This is because the natural extension of statistical mechanics to non-Hermitian systems adopts the biorthogonal rather than the non-biorthogonal framework. The Loschmidt amplitude formally corresponds to \cite{supp}
\begin{align}
    \mathscr{G}_{k}(t)=\cos\left[h_{1}(k)t\right]-i\sin\left[h_{1}(k)t\right]\braket{P}_{0},
\end{align}
where we have denoted $\braket{P}_{0}=\tilde{p}_{k}^{+}\braket{\tilde{u}_{+}^{0}(k)|\hat{\mathcal{H}}_{1}(k)|\tilde{u}_{+}^{0}(k)}+\tilde{p}_{k}^{-}\braket{\tilde{u}_{-}^{0}(k)|\hat{\mathcal{H}}_{1}(k)|\tilde{u}_{-}^{0}(k)}$, $\hat{\mathcal{H}}_{1}(k)=\mathcal{H}_{1}(k)/h_{1}(k)$, and $\tilde{p}_{k}^{\pm}=p_{k}^{\pm}/(p_{k}^{+}+p_{k}^{-})$.
Here, $\ket{\tilde{u}_{\pm}^{0}(k)}$ are the self-norm right eigenstates of $\mathcal{H}_{0}(k)$, satisfying $\braket{\tilde{u}_{\pm}^{0}(k)|\tilde{u}_{\pm}^{0}(k)}=1$, which is inconsistent with the biorthogonal relations. A similar expression for the rate function $\lambda'(t)$ can be derived within the non-biorthogonal formulation.

Following the derivation in the biorthogonal formulation, we can obtain the non-biorthogonal Fisher zeros (for integer $n$)
\begin{align}
    z'_{n}=\frac{i\pi(2n+1)}{2 h_{1}(k)}+\frac{1}{h_{1}(k)}\tanh^{-1}\braket{P}_{0}.
\end{align}
Again, by introducing the notation $\mathcal{Q}'_{k}=\tanh^{-1}\braket{P}_{0}$, $a=\pi(n+1/2)$, and decomposing 
$h_{1}(k)$ and $\mathcal{Q}'_{k}$ into their real and imaginary parts as $h_{1r}=\mathrm{Re}\left[h_{1}(k)\right],h_{1i}=\mathrm{Im}\left[h_{1}(k)\right],\mathcal{Q}'_{kr}=\mathrm{Re}\left[\mathcal{Q}'_{k}\right], \mathcal{Q}'_{ki}=\mathrm{Im}\left[\mathcal{Q}'_{k}\right]$, we obtain the non-biorthogonal critical-$k_c$ condition $a h_{1i}+h_{1r}\mathcal{Q}'_{kr}+h_{1i}\mathcal{Q}'_{ki}=0$.
It implies that the two vectors $\vec{v}'_{s}=(\mathcal{Q}'_{kr},\mathcal{Q}'_{ki}+a)$ and $\vec{v}_{h}=(h_{1r},h_{1i})$ in the 2D plane are orthogonal, i.e.,
\begin{align}
    \label{eqnoncrticalcon}
    \vec{v}'_{s}\cdot \vec{v}_{h}=0.
\end{align}
The ensuing geometric and topological interpretations closely parallel those in the biorthogonal framework. Particularly, one can define a winding number relation in the non-biorthogonal framework analogous to Eq. (\ref{eqwindiff}), where the two formulations become equivalent in the Hermitian limit.

\emph{Topological characteristics of DQPTs with chiral symmetry.---} 
In the non-biorthogonal formulation, a distinctive winding topology arises during a quench between two chiral symmetric Hamiltonians of the form
\begin{align}
    \label{eqtwochiral}
    \mathcal{H}_{s}(k)=d_{x}\sigma_{x}+d_{y}\sigma_{y},
\end{align}
specifically from a non-Hermitian initial Hamiltonian $\mathcal{H}_{s,0}(k)$ to a Hermitian final Hamiltonian $\mathcal{H}_{s,1}(k)$, under ground-state-like participation probabilities $p_{k}^{-} = 1$ and $p_{k}^{+} = 0$. Applying the sufficient condition of DQPTs given in Eq. (\ref{eqnoncrticalcon}), where the postquench energy $d_{1}(k)=\left(d_{x,1}^{2}+d_{y,1}^{2}\right)^{1/2}$ and $\braket{P}_{0}=\braket{\tilde{u}_{-}^{0}(k)|\hat{\mathcal{H}}_{s,1}(k)|\tilde{u}_{-}^{0}(k)}$ are both real value, we obtain that $\braket{P}_{0}=0$ implies the occurrence of the DQPTs. After some algebra, the critical-$k_c$ condition is determined by the orthogonality between two vectors in the complex plane---the prequench vector $\mathcal{R} e^{i\phi^{0}}$ and the normalized postquench vector $\hat{d}_{1}=\hat{d}_{x,1}+i\hat{d}_{y,1}$ with the angle $\phi^{1}$, where the real-value $\mathcal{R}$ is computed from the prequench Hamiltonian (see Supplementary Materials \cite{supp} for details). In other word, when the angle difference $\Delta\phi=\phi^{1}-\phi^{0}$ covers $m\pi/2$ (with odd $m$), the DQPTs must occur. Therefore, the nonzero winding number difference
\begin{align}
    \label{eqwindingrela}
    \Delta \nu=\nu^{1}-\nu^{0}=\mathbb{Z}/2,
\end{align}
is the sufficient condition of the occurrence of DQPTs, where
\begin{align}
    \label{eqnonwinding}
    \nu^{0}&=\frac{1}{2\pi}\oint_{\mathrm{BZ}} dk \, \frac{d_{x,0} \partial_{k} d_{y,0} -d_{y,0} \partial_{k} d_{x,0}}{d_{x,0}^2+d_{y,0}^2}, \\
    \nu^{1}&=\frac{1}{2\pi i}\oint_{\mathrm{BZ}} dk \, \hat{d}_{1}^{-1}\partial_{k}\hat{d}_{1},
\end{align}
and $\nu^{0}$ ($\nu^{1}$) is the half-integer (integer) quantized non-Hermitian (Hermitian) winding number \cite{supp}. Interestingly, since the underlying equilibrium topological phases of the non-Hermitian prequench and Hermitian postquench Hamiltonians are characterized by these two winding numbers, which indicate the existence of edge states \cite{yin2018geometric}, such DQPTs can be identified as a new type of dynamical topological phase transition. Beyond the above zero-temperature ground-state case, the same topological characteristic of DQPTs is conditionally established in finite-temperature scenario, as detailed in the Supplementary Materials \cite{supp}.

\begin{figure}
    \centering
     \includegraphics[scale=0.21]{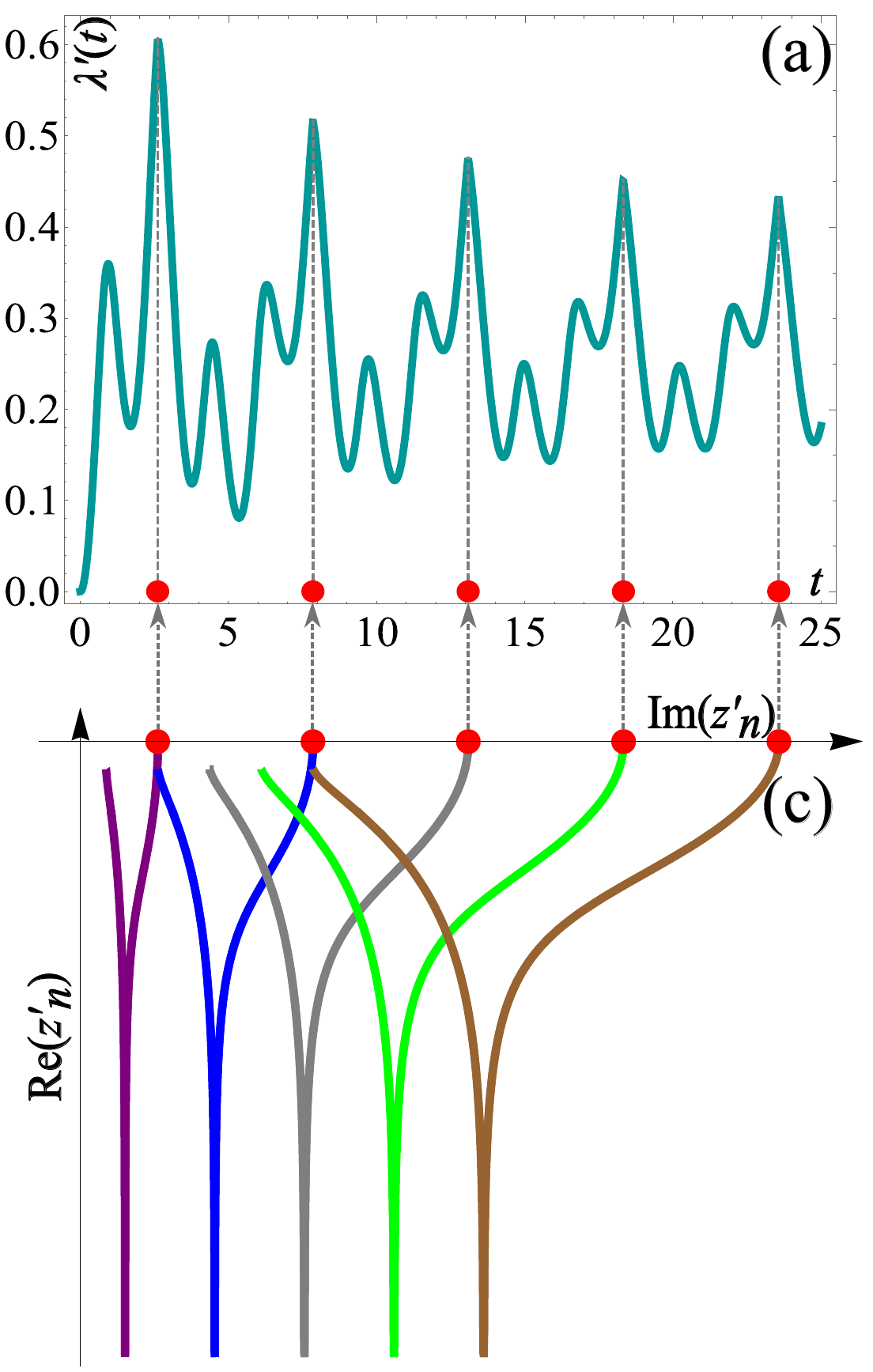}
     \includegraphics[scale=0.212]{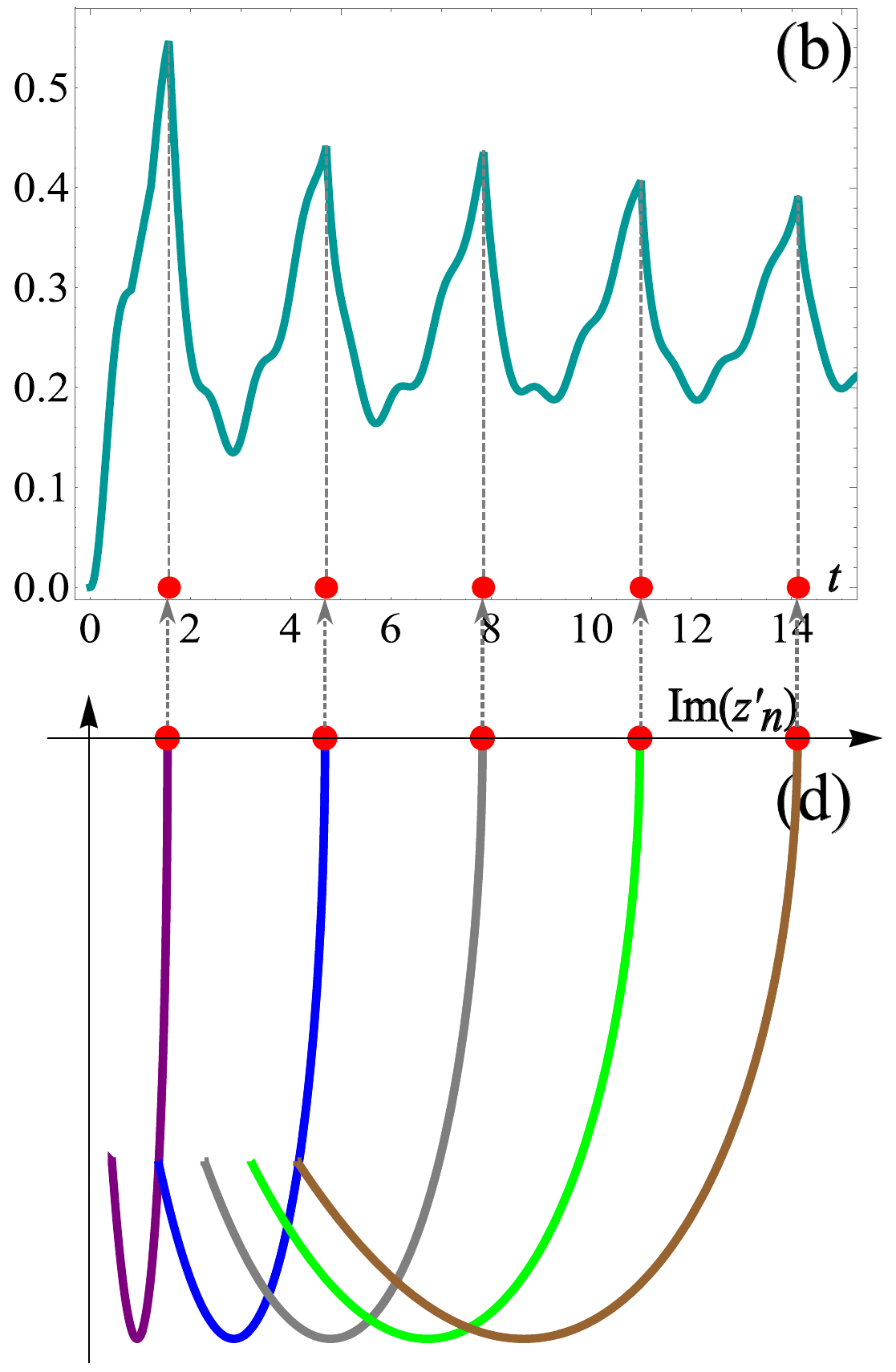}\\
     \includegraphics[scale=0.235]{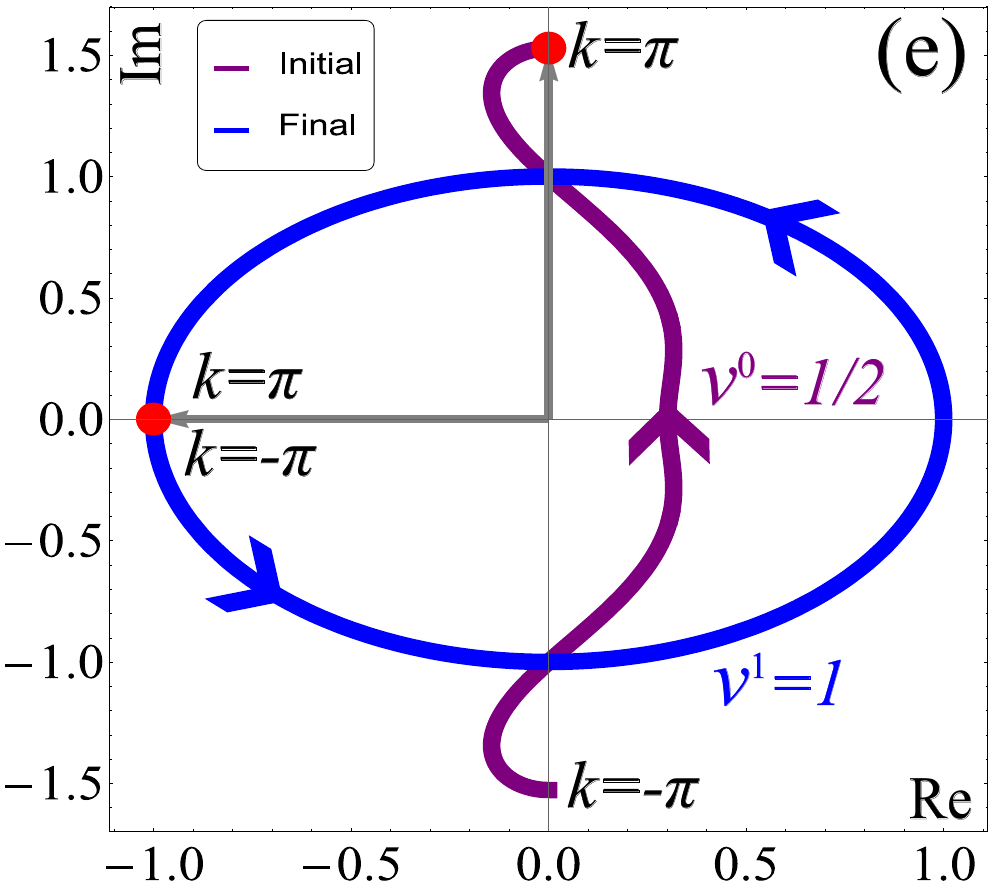}
     \includegraphics[scale=0.235]{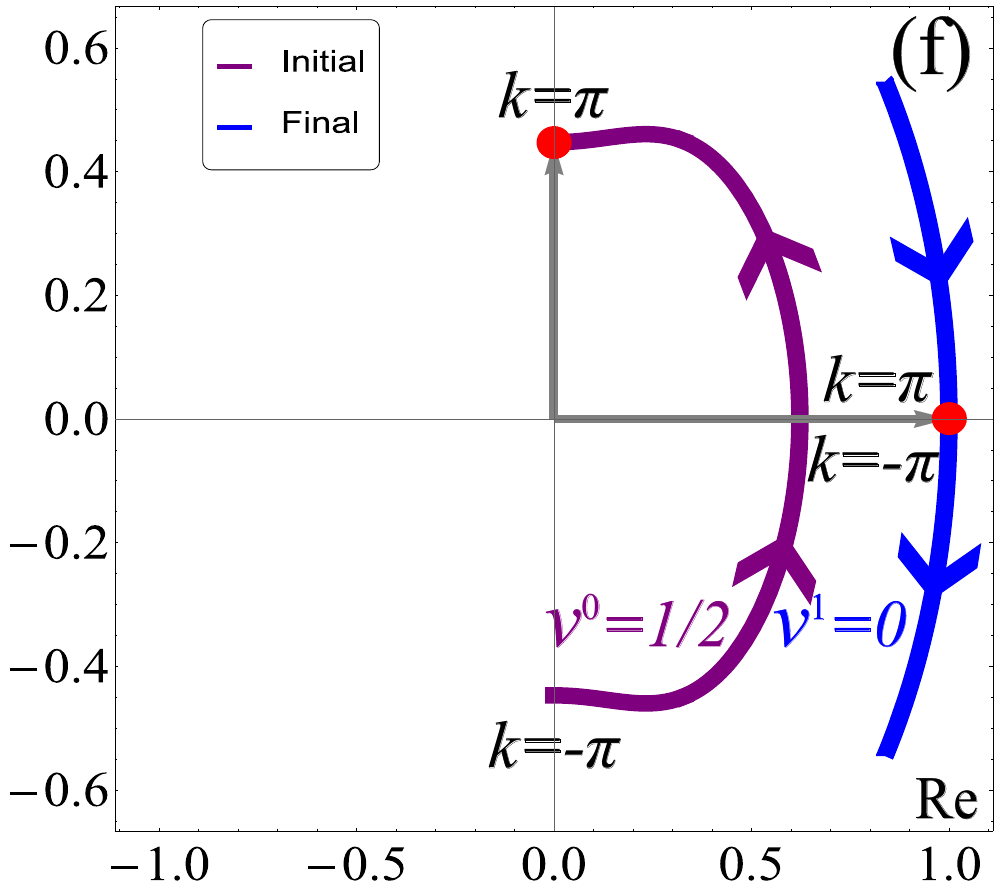}
     \caption{The topological characteristic of the non-Hermitian SSH model under a quantum quench from $\gamma=1.5$ to $0$, are analyzed with fixed parameters $t_{1}=0.6$ [panels (a), (c), and (e)] or $2.2$ [panels (b), (d), and (f)], and $t_{2}=1$. (a) and (b) The rate function $\lambda'(t)$ as a function of time $t$, with the DQPTs indicated by the non-analytical cusps marked by red points along the $t$ axis. These critical times correspond precisely to the intersections of Fisher zero flows with the imaginary axis in the complex plane, as illustrated in panels (c) and (d) by the Fisher zeros $z'_{n}$ ($n=0,1,2,3,4$) arranged from left to right. Panels (e) and (f) depict the momentum-resolved flows of the vectors $\mathcal{R} e^{i\phi^{0}}$ and $\hat{d}_{1}$ associated with the initial and final Hamiltonians. As $k$ varies in $[-\pi, \pi]$, the winding number differences $\Delta \nu=1/2$ and $-1/2$ emerge. Notably, the blue flow in (f) evolves from $-\pi$, descends to the lower endpoint, then ascends to the upper endpoint, and finally returns downward to $\pi$ (identical to $-\pi$), resulting in a vanishing winding number. Red pints in (e) and (f) mark the critical momenta $k_c$, where the two vectors are orthogonal (as indicated by the gray arrows), signaling the onset of DQPTs.}
    \label{figtopo}
\end{figure}

To exemplify the topological characteristic of DQPTs in the basic zero-temperature scenario, we consider the chiral symmetric non-Hermitian Su-Schrieffer-Heeger (SSH) model with $d_{x}=t_1+t_2\cos k$ and $d_{y}=t_2\sin k+i\gamma$. The quantum quench scenario is set via $\gamma$ from nonzero (non-Hermitian, e.g., $1.5$) to zero (Hermitian), with other parameters fixed as $t_{1}=0.6$ [Figs. \ref{figtopo} (a), (c), and (e)] or $t_{1}=2.2$ [Figs. \ref{figtopo} (b), (d), and (f)], and $t_{2}=1$. The intersection points between the Fisher zero flows and the imaginary axis, as momentum $k$ varies, indeed encode the critical times of the DQPTs, which are marked by the non-analytical points (cusps) in the (time-evolved normalized) rate function, as illustrated in Figs. \ref{figtopo}(a)-(d). Importantly, the emergence of the DQPTs is characterized by the topological winding numbers between initial (prequench) and final (postquench) Hamiltonians, as shown in Figs. \ref{figtopo}(e) and (f), with $\Delta \nu=1/2$ and $-1/2$, respectively, where the final Hamiltonians are topologically nontrivial ($\nu^{1}=1$) or trivial ($\nu^{1}=0$) while the initial Hamiltonians are both topological ($\nu^{0}=1/2$). Specifically, the critical $k_{c}=\pi$ is signaled by the orthogonality between the two vectors $\mathcal{R} e^{i\phi^{0}}$ and $\hat{d}_{1}$.

From the quantum physics perspective, open quantum systems provide an important avenue for realizing non-Hermitian systems \cite{song2019damping, liu2020helical, xue2022}. In our setup, a Hermitian SSH model with flux-controlled dissipation, described by the Lindblad master equation, exhibits a non-unitary short-time evolution of a non-biorthogonal density matrix that is effectively governed by a non-Hermitian SSH Hamiltonian prior to the occurrence of quantum jumps (see Supplementary Materials \cite{supp} for details). The quantum quench is achieved by abruptly turning off such dissipation, thereby driving a transition from a non-Hermitian SSH model to its Hermitian counterpart. Furthermore, the long-time non-unitary evolutions, anticipated to further substantiate our DQPT framework beyond the short temporal window, can be captured via the dynamical equations of single-particle correlation functions \cite{song2019damping} or the mean-field treatment of annihilation operators \cite{zhou2025floquet}. In standard experimental implementations, our framework is expected to be validated via photonic quantum walks, building on relevant studies \cite{wang2019simu, zhang2025self}. Moreover, electrical circuit platforms can simulate topological and non-Hermitian systems \cite{lee2018topo, helbig2020, hu2021knots}, including the non-Hermitian SSH model where the non-reciprocal hoppings can be implemented using the impedance converter with current inversion \cite{supp}. Nevertheless, realizing quantum time-evolution simulations in electrical circuits remains a significant challenge, owing to the complexity of simulating the time-dependent Schrödinger equation \cite{schindler2011exp, zhang2023electrical, zhou2025floquet}, and requires further exploration.

\emph{Conclusions.---} In this work, we have systematically analyzed non-Hermitian DQPTs from both biorthogonal and non-biorthogonal perspectives, extending the formulation to encompass mixed-state dynamics. Our unified treatment of Loschmidt amplitudes and echos captures the essential geometric features of DQPTs in the two-band model. Furthermore, the discovery of topological signatures in a special non-biorthogonal quench scenario---establishing an analytical relationship between the equilibrium topological phases of the prequench and postquench systems---indicates that the topological characteristics of DQPTs transcends conventional Hermitian limit. These findings open avenues for further exploration of dynamical topology in many-body and experimentally relevant non-Hermitian quantum systems. In light of recent studies on the higher-dimensional non-Hermitian systems \cite{wang2024amoeba, chen024hybrid, xie2025dirac, shi2025floquet}, our theoretical framework can be combined with these insights to further guide the exploration of DQPTs in such higher-dimensional non-Hermitian systems.

\emph{Acknowledgment.---} We thank Haifeng Lang and Yang Zhou for insightful discussions. Y.F. is supported by a startup grant from Zhejiang Normal University. G.X. acknowledges support from the NSFC under Grant No. 12174346.

\bibliography{ref}

\newpage

\begin{widetext}

\begin{center}
    \textbf{\large Supplementary Materials for ``Anatomy of Non-Hermitian Dynamical Quantum Phase Transitions''}
\end{center}

\section{Parallel transport and Loschmidt amplitudes in Hermitian systems}
In this paper, we aim to develop a comprehensive mixed-state formulation for dynamical quantum phase transitions (DQPTs), grounded in the most fundamental mathematical and physical principles. In Hermitian systems, this necessarily involves the geometric phases of the mixed-state density matrices, which are closely related to the concept of the parallel transport \cite{pancharatnam1956,samuel1988,cohen2019,simon1983,berryphase1984,aharonov1987,uhlmann1986,uhlmann1989,uhlmann1991,hubner1992,hubner1993,ajoqvist2000,viyuela2014,bhattacharya2017mixed}. Parallel transport can be defined on various structures in mathematics. For instance, on a manifold equipped with the affine connection, the parallel transport of a vector field along a curve is defined by the condition that the covariant derivative of the vector field along the curve's tangent vector vanishes \cite{nakahara2003book,Stone_Goldbart_2009}. However, the situation is more intricate in the context of the DQPTs we are currently investigating, as it essentially entails a parallel transport problem on a principal bundle \cite{uhlmann1986,uhlmann1989,uhlmann1991,hubner1993, budich2015topo}.
Formally, a fibre bundle consists of three differentiable manifolds: the total space $\mathbb{E}$, the base space $\mathbb{M}$, and the fibre $\mathbb{F}$. It is equipped with a surjection $p: \mathbb{E} \rightarrow \mathbb{M}$, a Lie group $\mathbb{G}$ (known as the structure group) that acts on $\mathbb{F}$, a set of open covering $\left\{U_i\right\}$ that provide the coordinates for $\mathbb{M}$, and the transition functions that define the transitions between any two overlapping charts $U_i \cap U_j\neq \varnothing$ (for a detailed introduction, see Ref. \cite{nakahara2003book} or other relevant references). A principal bundle has a fibre $\mathbb{F}$ which is identical to the structure group $\mathbb{G}$.  

Consider a curve $\gamma: [0,1]\rightarrow \mathbb{M}$ on the base space, a curve $\tilde{\gamma}: [0,1] \rightarrow \mathbb{E}$ on the total space of a principal bundle is said to be a horizontal lift of $\gamma$ if $p \circ \tilde{\gamma}=\gamma$ and the tangent vector to $\tilde{\gamma}(t)$ always belongs to the horizontal subspace $\mathbb{H}_{\tilde{\gamma}(t)}\mathbb{E}$ \cite{nakahara2003book}. Any point on $\tilde{\gamma}(t)$ is defined as the parallel transport of preceding points on $\tilde{\gamma}(t)$. Here, we involve the abstract definition for the connection on a principal bundle that the a connection on $\mathbb{E}$ provides a unique separation of the tangent space $\mathbb{T}_{u}\mathbb{E}$ into the vertical subspace $\mathbb{V}_{u}\mathbb{E}$ and the horizontal subspace $\mathbb{H}_{u}\mathbb{E}$ for arbitrary point $u$ on $\mathbb{E}$. Visually, the inverse imagine of $p$ at a point $b$ on $\mathbb{M}$ is a fibre on $\mathbb{E}$, which is isomorphic to the structure group $\mathbb{G}$ and can be thought of as a strand of hair extending above $b$. The vertical subspace $\mathbb{V}_{u}\mathbb{E}$ can be roughly interpreted as the tangent vector along the fibre $\mathbb{F}$ (equivalent to $\mathbb{G}$), while the horizontal subspace $\mathbb{H}_{u}\mathbb{E}$ is orthogonal to $\mathbb{F}$ [Fig. \ref{supfig-bundle}(a)]. This parallel transport is visualized as the movement of a curve perpendicular to the fiber. In other words, the tangent vector along $\tilde{\gamma}(t)$ has no projection onto the tangent space of $\mathbb{F}$ at any point.    

\begin{figure}
    \centering
    \includegraphics[width=0.8\linewidth]{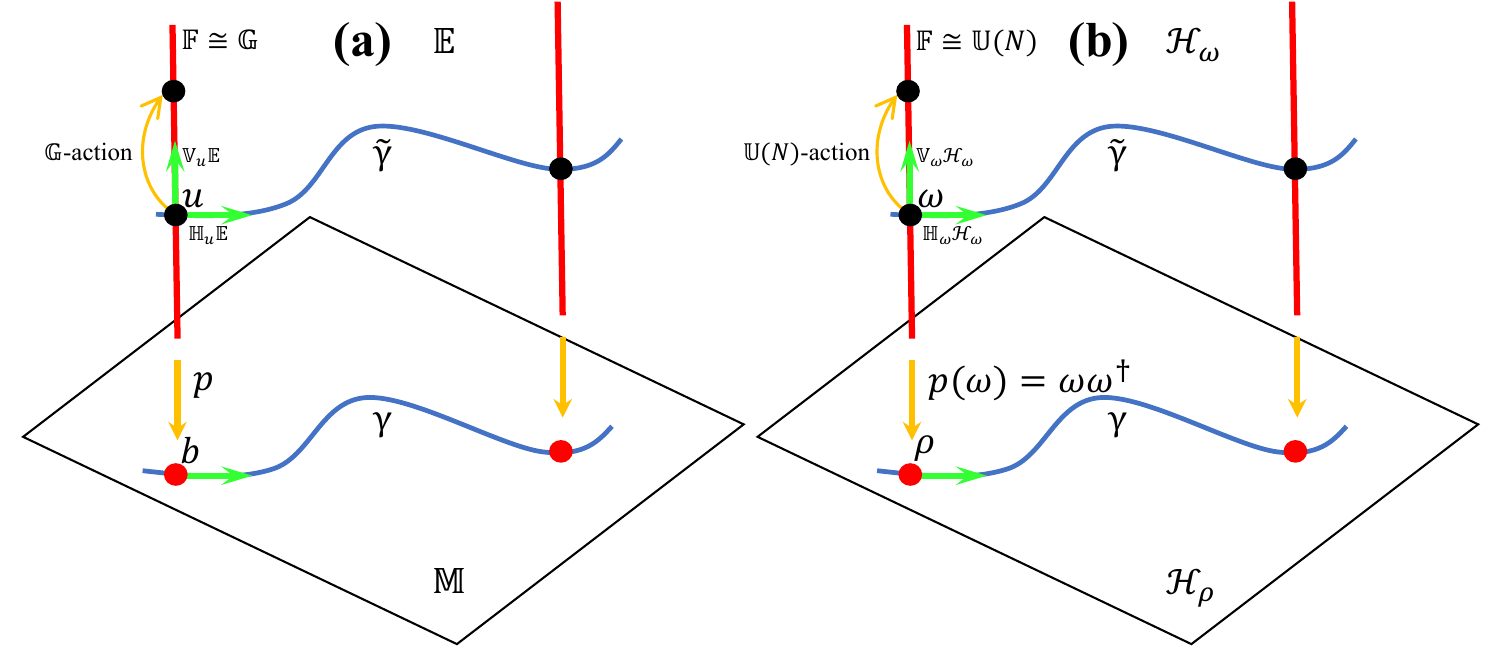}
    \caption{The intuitive schematic of parallel transport on principal bundles. (a) This illustrates the total space $\mathbb{E}$, the base space $\mathbb{M}$, and the fibre $\mathbb{F}$, which is equivalent to the structure group $\mathbb{G}$. A point $u$ on a fibre is projected to the point $b$ on $\mathbb{M}$ via the surjection $p$. A curve $\gamma$ in $\mathbb{M}$ is horizontal lifted to $\tilde{\gamma}$ in $\mathbb{E}$. The tangent vector at $u$ along $\tilde{\gamma}$ belongs to the horizontal subspace $\mathbb{H}_{u}\mathbb{E}$, which projects to the tangent vector at $b$ along $\gamma$. $\mathbb{H}_{u}\mathbb{E}$ is visually orthogonal to the vertical subspace $\mathbb{V}_{u}\mathbb{E}$ along $\mathbb{F}$. The structure group $\mathbb{G}$ acts on $u$ from the right, mapping it to another point on the same fibre. (b) A physical realization of parallel transport in the Hilbert space of the density matrix $\rho$ and its amplitude $\omega$, with the structure group $\mathbb{U}(N)$.}
    \label{supfig-bundle}
\end{figure}

\subsection{Parallel transport and geometric phases in Hermitian systems}

Returning to the physical context about the geometric phases of mixed-state density matrices, we define the base space $\mathcal{H}_{\rho}$ as the space associated with the (nonnegative trace class) density matrix $\rho$ of a Hermitian system governed by the Hamiltonian $\hat{H}$. The amplitude $\omega$ of the density matrix $\rho$ is then defined by,
\begin{align}
    \label{eqamplitudedef}
    \rho=\omega\omega^{\dagger},
\end{align}
which essentially specifies the surjection $p$ from the total space, identified as the Hilbert space $\mathcal{H}_{\omega}$ of the amplitude [Fig. \ref{supfig-bundle} (b)], to the base space $\mathcal{H}_{\rho}$ \cite{uhlmann1986,uhlmann1989,uhlmann1991,hubner1992,hubner1993,viyuela2014,bhattacharya2017mixed}. The explicit form of the amplitude is given by
\begin{align}
    \omega=\sqrt{\rho}U,
\end{align}
where $U$ is an arbitrary unitary matrix representing the gauge freedom. This implies that the structure group is the unitary group $\mathbb{U}(N)$, where $N$ corresponds to the degrees of freedom of the Hilbert space. Mathematically, the total space $\mathcal{H}_{\omega}$ can be rigorously identified with the complex general linear group $\mathrm{GL}(\mathbb{C},N)$, while the base space $\mathcal{H}_{\rho}$  is constructed as the quotient manifold $\mathrm{GL}(\mathbb{C},N)/\mathbb{U}(N)$. This quotient structure explicitly corresponds to the space of positive definite matrices \cite{budich2015topo}, where the $\mathbb{U}(N)$ symmetry is modded out through the equivalence relation $\omega\sim\omega U$ for $U\in \mathbb{U}(N)$. Despite the trivial topology of the principal bundle induced by the direct product decomposition $\mathrm{GL}(\mathbb{C},N)=\mathrm{GL}(\mathbb{C},N)/\mathbb{U}(N) \times \mathbb{U}(N)$, which ensures the existence of a global gauge \cite{budich2015topo}, the framework of this principal bundle remains fully capable of accommodating parallel transport, which is achieved through the construction outlined in the preceding discussion.

However, the abstract concept of parallel transport discussed above is somewhat difficult to apply directly. Uhlmann, however, presents a geometric framework that facilitates the implementation of parallel transport \cite{uhlmann1986, uhlmann1989, uhlmann1991}. To transport a given $\omega_{1}$ from the total space, associated with $\rho_{1}=p(\omega_{1})$ to the fibre over $\rho_{2}$, one simply selects the unique $\omega_{2}$ that minimizes the distance among the possible options \cite{hubner1993}. Here, Uhlmann's transport parallel transport is based on the Hilbert-Schmidt metric (distance) 
\begin{align}
    \label{suppeqdis}
    d(\omega_{1},\omega_{2})^{2}=\mathrm{Tr}(\omega_{1}-\omega_{2})(\omega_{1}-\omega_{2})^{\dagger},
\end{align}
which induces the parallel transport condition \cite{uhlmann1986, uhlmann1989, uhlmann1991, hubner1992, hubner1993}
\begin{align}
\label{condition1}
    \omega_{1}^{\dagger}\omega_{2}=\omega_{2}^{\dagger}\omega_{1}>0.
\end{align}
The standard infinitesimal parallel transport on principal bundles can be generalized to a form of distant parallel transport along a sequence of base points. In differentiable settings, the conventional concepts are regained by taking the appropriate limits \cite{hubner1993}. To this end, assume that $\omega_{1}$ and $\omega_{2}$ are infinitesimally close, i.e., $\omega_{2}=\omega_{1}+\dot{\omega}_{1}ds$ with the curve parameter $s$, and we drop the subscript for simplicity. We decompose the tangent vector $\dot{\omega}$ into its vertical and horizontal components $\dot{\omega}=\dot{\omega}_{V}+\dot{\omega}_{H}$, where $\dot{\omega}_{V}\in\mathbb{V}_{\omega}\mathcal{H}_{\omega}$ and $\dot{\omega}_{H}\in\mathbb{V}_{\omega}\mathcal{H}_{\omega}$. Here, $\dot{\omega}$ explicitly denotes a tangent vector $\dot{\omega}^{\mu}\partial/\partial x^{\mu}$ at the point $\omega$, where $\dot{\omega}^{\mu}=dx^{\mu}(\omega)/ds$ is the component in the local coordinate $\left\{x^{\mu}\right\}$. The vertical component $\dot{\omega}_{V}$ 
  is defined such that for an arbitrary smooth function $f: \mathcal{H}_{\omega}\rightarrow \mathbb{R}$, the following holds \cite{nakahara2003book}
\begin{align}
    \dot{\omega}_{V}f(\omega)=\frac{d}{ds}f\left[\omega \exp(s A)\right]\Big|_{s=0},
\end{align}
where $A=-A^{\dagger}$ is a generator of $\mathbb{U}(N)$. Expanding this in the coordinate representation, we obtain
\begin{align}
    \dot{\omega}_{V}^{\mu}\frac{\partial f(\omega)}{\partial x^{\mu}}=\frac{\partial f\left[\omega \exp(s A)\right]}{\partial x^{\mu}} \left[\omega A \exp(s A)\right]^{\mu}\Big|_{s=0}= (\omega A)^{\mu}\frac{\partial f(\omega)}{\partial x^{\mu}},
\end{align}
which implies that $ \dot{\omega}_{V}=\omega A$. The horizontal component $\dot{\omega}_{H}$ is orthogonal to $\dot{\omega}_{V}$, i.e., $\mathrm{Tr} \left[\dot{\omega}_{H}^{\dagger}\omega A\right]=0$. Combing this with its Hermitian conjugate yields $\mathrm{Tr} \left[A(\dot{\omega}_{H}^{\dagger}\omega -\omega^{\dagger}\dot{\omega}_{H})\right]=0$. This condition is precisely the Uhlmann parallel transport condition
\begin{align}
    \label{condition2}
    \dot{\omega}^{\dagger}\omega -\omega^{\dagger}\dot{\omega}=0,
\end{align}
which must be satisfied by the horizontal tangent vector $\dot{\omega}$ \cite{uhlmann1986, uhlmann1989, uhlmann1991, budich2015topo}. This condition represents the infinitesimal form of the condition given in Eq. (\ref{condition1}).

Beyond the mathematical framework, parallel transport can also be defined within the experimental context of quantum interferometry \cite{ajoqvist2000}. It is well-known that the Hermitian density matrix evolves unitarily with time, expressed as $\rho(t) = U(t) \rho(0) U(t)^{\dagger}$, where $\rho_0$ is the initial state and $U(t) = e^{-i\hat{H}t}$. Parallel transporting a mixed state along an arbitrary path means that at each instant, the state
must be in phase with the state at an infinitesimal time. This condition leads to the physical parallel transport condition \cite{ajoqvist2000}
\begin{align} 
\mathrm{Tr}\left[\rho(t)\dot{U}(t)U(t)^{\dagger}\right] = 0. 
\end{align}
Such parallel transport imposes the vanishing of the dynamical phase at any time $t$, 
\begin{align}
    \label{supeqdyn}
    \phi_{dyn}=-\int_{0}^{t}dt'\mathrm{Tr}\left[\rho(t')\hat{H}\right]=-i\int_{0}^{t}dt'\mathrm{Tr}\left[\rho(0)U^{\dagger}(t')\dot{U}(t')\right]=0,
\end{align}
where we have used 
\begin{align}
    \mathrm{Tr}\left[\rho(0)U^{\dagger}(t')\dot{U}(t')\right]&=\mathrm{Tr}\left[U(t')\rho(0)U^{\dagger}(t')\dot{U}(t')U^{\dagger}(t')\right]\nonumber\\
    &=\mathrm{Tr}\left[\rho(t')\dot{U}(t')U^{\dagger}(t')\right]\nonumber\\
    &=\mathrm{Tr}\left[\rho(t')(-i\hat{H})U(t')U^{\dagger}(t')\right]\nonumber\\
    &=-i\mathrm{Tr}\left[\rho(t')\hat{H}\right].
\end{align}

The parallel transport evolution $V(t)$ is uniquely generated by the connection $\mathscr{A}$ on the principal bundles \cite{nakahara2003book}
\begin{align}
    V(t)=\mathcal{P}e^{\int_{0}^{t} dt' \mathscr{A}(t')},
\end{align}
which results in an group element of $\mathbb{U}(N)$ acting on the fibre and $\omega(t)=\omega(0)V(t)$. If we define $\sqrt{\rho(t)}$ as a section from $\mathcal{H}_{\rho}$ to $\mathcal{H}_{\omega}$, this implies that the initial gauge is given by $\omega(0) = \sqrt{\rho(0)} U_{g}(0)$, where $U_{g}(0) = 1$. Consequently, the parallel transport specifies $\omega(t) = \sqrt{\rho(t)} V(t)$. Any initial gauge choice $U_{g}(0)$ specifies a section $\sqrt{\rho(t)}U_{g}(t)$ and leads to $\omega(t)=\sqrt{\rho(t)}V(t)U_{g}(0)$, that is $\omega(t)=\sqrt{\rho(t)}U_{g}(t)V_{g}(t)$, where $V_{g}(t)=U_{g}^{\dagger}(t)V(t)U_{g}(0)$. This gauge transformation of the parallel transport is illustrated in Fig. \ref{supfig-gauge}. The holonomy group, the subgroup of $\mathbb{U}(N)$, is constituted by the collection of $V(t)$ along some closed parallel transport paths \cite{nakahara2003book}. Even though the path is not closed, the phase 
\begin{align}
\label{supeqgeo}
    \phi_{g}=\arg \mathrm{Tr}\left[\omega(0)^{\dagger}\omega(t)\right],
\end{align}
remains gauge invariant. This phase represents a purely $\mathbb{U}(N)$-invariant geometric phase along any parallel transport path and corresponds to the Uhlmann phase in the case of a closed path \cite{uhlmann1986,uhlmann1989,uhlmann1991,viyuela2014,bhattacharya2017mixed}. The Uhlmann phase serves as a natural extension of the renowned Berry phase, applicable to the mixed-state scenario \cite{simon1983,berryphase1984}.

Building on the considerations above, the Loschmidt amplitude for mixed states is defined as \cite{heyl2017mixed,bhattacharya2017mixed}
\begin{align}
    \label{suppeqlsa}
    \mathcal{G}(t)=\mathrm{Tr}\left[\rho(0)U(t)\right],
\end{align}
where $\rho(0)$ represents the density matrix of the prequench Hamiltonian $\hat{H}_{0}$, and $U(t)$ is the time evolution operator of the postquench Hamiltonian $\hat{H}_{1}$. This expression naturally reduces to the pure-state form $\mathcal{G}(t) = \braket{\Psi(0) | \Psi(t)}$, with the initial pure-state density matrix $\rho(0) = \ket{\Psi(0)}\bra{\Psi(0)}$. Note that the time evolution operator $U(t)$ does not necessarily correspond to a parallel transport path. More explicitly, Eq. (\ref{suppeqlsa}) can be rewritten as
\begin{align}
\mathcal{G}(t)&=\mathrm{Tr}\left[\sqrt{\rho(0)}^{\dagger} U(t) \sqrt{\rho(0)}\right]\nonumber\\
&=\mathrm{Tr}\left[\sqrt{\rho(0)} U(t) \sqrt{\rho(0)} U(t)^{\dagger} U(t)\right] \nonumber\\
&=\mathrm{Tr}\left[\sqrt{\rho(0)} \sqrt{\rho(t)}  U(t)\right],
\end{align}
which is equivalent to $\mathbb{U}(N)$-gauge invariant $\mathrm{Tr}\left[\omega(0)^{\dagger}\omega(t)\right]$ if $U(t) = V(t)$ corresponds to parallel transport with initial gauge $U_{g}(0)=1$. In this case, the geometric phase $\phi_{g}$ is given by $\phi_{g} = \arg \mathcal{G}(t)$. This relation holds for parallel transport with arbitrary gauge, $\mathcal{G}(t)=\mathrm{Tr}\left[U_{g}(0) \omega^{\dagger}(0) \sqrt{\rho(t)} V(t) U_{g}(0) U_{g}(0)^{\dagger}\right]=\mathrm{Tr}\left[\omega(0)^{\dagger}\omega(t)\right]$.
For a generic time evolution operator $U(t)$, however, an additional dynamical phase, as expressed in Eq. (\ref{supeqdyn}), arises due to the deviation from parallel transport. In the context of DQPTs, we relax the requirement to a $\mathbb{U}(1)$-invariant geometric phase, while $\mathbb{U}(N)$ invariance is not strictly necessary. To extract the such geometric phase, this dynamical phase must be subtracted
\begin{align}
\phi = \arg \mathcal{G} - \phi_{dyn},
\end{align}
ensuring gauge invariance under $\mathbb{U}(1)$ transformations \cite{bhattacharya2017mixed}. In the special case of a pure state, the geometric phase simplifies to the well-known Pancharatnam phase \cite{pancharatnam1956}.

\begin{figure}
    \centering
    \includegraphics[width=0.6\linewidth]{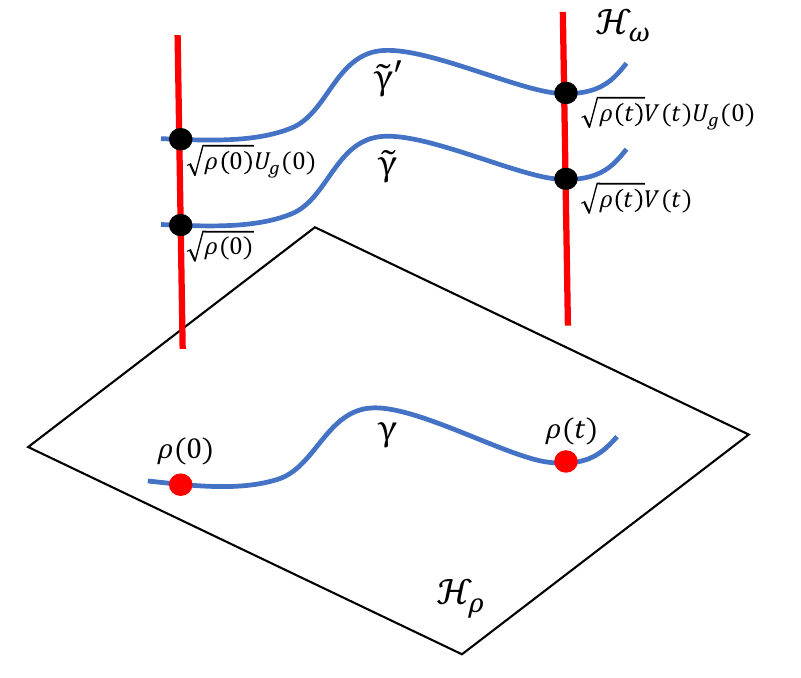}
    \caption{The visualization of gauge transformation for parallel transport on the principal bundle $\mathcal{H}_{\omega}\rightarrow \mathcal{H}_{\rho}$, where $\tilde{\gamma}'=\tilde{\gamma}U_{g}(0)$ and $\sqrt{\rho(t)}$ is a global section.}
    \label{supfig-gauge}
\end{figure}

\subsection{Purification formulation}

The parallel transport and geometric phase described above can be rephrased in terms of their purification formulation \cite{uhlmann1991, ajoqvist2000,viyuela2014}. To purify mixed-state density matrices, it suffices to consider an extended Hilbert space for the Hamiltonian system of the form
\begin{align} 
\mathscr{H}^{ext} = \mathscr{H}_{s} \otimes \mathscr{H}', 
\end{align} 
where the dimension of $\mathscr{H}'$ is at least as large as the dimension of $\mathscr{H}_{s}$. For clarity, let us assume a basis $\left\{\ket{\psi_{i}}\right\}$ for the Hilbert space $\mathscr{H}_{s}$. The density matrix can then be written as
\begin{align}
    \rho=\sum_{i}p_{i}\ket{\psi_{i}}\bra{\psi_{i}}, \quad p_{i}\geq 0
\end{align}
and the amplitude is expressed as
\begin{align}
    \omega=\sum_{i}\sqrt{p_{i}}\ket{\psi_{i}}\bra{\psi_{i}}U.
\end{align}
We define the pure state in the extended Hilbert space $\mathscr{H}^{ext}$ with the choice $\mathscr{H}' = \mathscr{H}_{s}$ as \cite{uhlmann1991,ajoqvist2000,viyuela2014}
\begin{align}
\label{supeqpurestate}
    \ket{\omega}=\sum_{i}\sqrt{p_{i}}\ket{\psi_{i}}\otimes U^{T}\ket{\psi_{i}},
\end{align}
where $U^{T}$ denotes the transpose of the unitary matrix $U$. The corresponding density operator is then given by
\begin{align}
    \rho=\mathrm{Tr}_{2}\left(\ket{\omega}\bra{\omega}\right),
\end{align}
where $\mathrm{Tr}_{2}$ denotes the partial trace over $\mathscr{H}'$. We observe the relation $\braket{\omega|\omega'}=\mathrm{Tr}\left[\omega\omega^{'\dagger}\right]$. The Loschmidt amplitude of the pure state in the extended Hilbert space $\mathscr{H}^{ext}$ is 
\begin{align}
    \mathcal{G}_{p}(t)=\braket{\omega(0)|\omega(t)},
\end{align}
where we assume the prequench state as Eq. (\ref{supeqpurestate}), and the postquench state is
\begin{align}
     \ket{\omega(t)}=\sum_{i}\sqrt{p_{i}}U(t)\ket{\psi_{i}}\otimes U^{T}\ket{\psi_{i}}.
\end{align}
Note that $U(t)$ only acts on the first state in the physical Hilbert space $\mathscr{H}_{s}$. It is explicit that
\begin{align}
    \mathcal{G}_{p}(t)&=\sum_{i,j}\left(\sqrt{p_{i}}\bra{\psi_{i}}\otimes \bra{\psi_{i}}U^{*}\right)\left(\sqrt{p_{j}}U(t)\ket{\psi_{j}}\otimes U^{T}\ket{\psi_{j}}\right)\nonumber\\
    &=\sum_{i}p_{i}\bra{\psi_{i}}U(t)\ket{\psi_{i}}\nonumber\\ &=\sum_{i,j}\braket{\psi_{i}|p_{j}\ket{\psi_{j}}\bra{\psi_{j}}U(t)|\psi_{i}}\nonumber\\
    &=\mathrm{Tr}\left[\rho(0)U(t)\right]=\mathcal{G}(t).
\end{align}
The dynamical phase can also be expressed as 
\begin{align}
    \phi_{dyn}=-\int_{0}^{t}dt'\braket{\omega(t')|\hat{H}_{1}|\omega(t')},
\end{align}   
which follows from the derivation
\begin{align}
 \phi_{dyn}&=\int_{0}^{t}dt'\sum_{m,n}\left(\sqrt{p_{m}}\bra{\psi_{m}}U(t')^{\dagger}\otimes \bra{\psi_{m}}U^{*}\right)\hat{H}_{1}\left(\sqrt{p_{n}}U(t')\ket{\psi_{n}}\otimes U^{T}\ket{\psi_{n}}\right)\nonumber\\
    &=-\int_{0}^{t}dt'\sum_{m}p_{m}\bra{\psi_{m}}U(t')^{\dagger}\hat{H}_{1}U(t')\ket{\psi_{m}} \nonumber\\
    &=-i\int_{0}^{t}dt'\sum_{m}p_{m}\bra{\psi_{m}}U(t')^{\dagger}\dot{U}(t')\ket{\psi_{m}} \nonumber\\
    &=-i\int_{0}^{t}dt'\sum_{m,n}p_{m}\braket{\psi_{m}|\psi_{n}}\bra{\psi_{n}}U(t')^{\dagger}\dot{U}(t')\ket{\psi_{m}} \nonumber\\
    &=-i\int_{0}^{t}dt'\sum_{m,n}p_{n}\braket{\psi_{m}|\psi_{n}}\bra{\psi_{n}}U(t')^{\dagger}\dot{U}(t')\ket{\psi_{m}} \nonumber\\
    &=-i\int_{0}^{t}dt'\sum_{m}\bra{\psi_{m}}\rho(0)U(t')^{\dagger}\dot{U}(t')\ket{\psi_{m}} \nonumber\\
    &=-\int_{0}^{t}dt'\mathrm{Tr}\left[\rho(t')\hat{H}_{1}\right].
\end{align}
Thus, we can fully reformulate the geometric phase and Loschmidt amplitude in terms of their purification formulation, where the parallel transport condition is given by \cite{ajoqvist2000} 
\begin{align} \braket{\omega(t)|\hat{H}_{1}|\omega(t)}=\braket{\omega(t)|\dot{\omega}(t)}=0, \end{align} 
which enforces the cancellation of the dynamical phase.

\section{Non-Hermitian dynamics}
\label{nhdynamic}

In (Hermitian) quantum mechanics, the time evolution is governed by the Schr\"odinger equation $i\partial_{t}\ket{\psi(t)}=\hat{H}\ket{\psi(t)}$ for the time-dependent state of the time-dependent state $\ket{\psi(t)}$ in the Schr\"odinger picture and Heisenberg equation $\partial_{t}\hat{O}(t)=-i\left[\hat{O}(t),\hat{H}\right]$ for the time-dependent operator $\hat{O}(t)$ in the Heisenberg picture, where $\ket{\psi(t)}=e^{-i\hat{H}t}\ket{\psi(0)}$ and $\hat{O}(t)=e^{i\hat{H}t}\hat{O}e^{-i\hat{H}t}$ with the unitary time-evolution operator $U(t)=e^{-i\hat{H}t}$.

Breaking the Hermiticity, the biorthogonal (stationary) eigenstates emerge for non-Hermitian Hamiltonian $\hat{H}\neq \hat{H}^{\dagger}$,
\begin{align}
  &\mathrm{right \,\, eignestate}\qquad  \hat{H}\ket{u_n}=\epsilon_n\ket{u_n},\\
  &\mathrm{left \, \,eignestate}\qquad \,\,   \hat{H}^{\dagger}| u_{n} \rangle\!\rangle=\epsilon^*_n| u_{n} \rangle\!\rangle.
\end{align}
Using the completeness relation $\sum_{n}\ket{u_{n}}\langle\!\langle u_{n}|=\hat{1}$, the time evolution of a state expressed in terms of the right eigenstates is
\begin{align}
    \ket{\psi(t)}=e^{-i\hat{H}t}\ket{\psi(0)}=\sum_{n}\mathscr{C}_{n}(t)\ket{u_{n}},
\end{align}
where $\mathscr{C}_{n}(t)=\langle\!\langle u_{n}|e^{-i\hat{H}t}|\psi(0)\rangle=\mathscr{C}_{n}e^{-i\epsilon_{n}t}$ with $\mathscr{C}_{n}=\langle\!\langle u_{n}|\psi(0)\rangle$. To compute observables and expectation values correctly, we also need to define the corresponding associate state $\bra{\tilde{\psi}(t)}$, which can be done in three different manners, as explained below.

In the first manner, we choose the Hermitian conjugate of $\ket{\psi(t)}$ such as 
\begin{align}
    \bra{\tilde{\psi}(t)}=\bra{\psi(t)}=\bra{\psi(0)}e^{i\hat{H}^{\dagger}t}=\sum_{n}\mathscr{C}_{n}^{*}(t)\bra{u_{n}},
\end{align}
where $\mathscr{C}_{n}^{*}(t)=\bra{\psi(0)}e^{i\hat{H}^{\dagger}t}|u_{n}\rangle\!\rangle = \mathscr{C}_{n}^{*}e^{i\epsilon^{*}t}$. That means the Heisenberg equation becomes $\partial_{t}\hat{O}(t)=-i\hat{O}(t)\hat{H}+i\hat{H}^{\dagger}\hat{O}(t)$, and the observable of $\hat{O}(t)$ is 
\begin{align}
    \langle \hat{O}(t) \rangle =\braket{\psi(t)|\hat{O}|\psi(t)}=\sum_{nm}\mathscr{C}_{n}^{*}(t)\mathscr{C}_{m}(t)\braket{u_{n}|\hat{O}|u_{m}}=\sum_{nm}\mathscr{C}_{n}^{*}\mathscr{C}_{m}e^{i(\epsilon_{n}^{*}-\epsilon_{m})t}\braket{u_{n}|\hat{O}|u_{m}}.
\end{align}
This expectation value requires normalization by $1/\braket{\psi(t)|\psi(t)}$ at each time due to the decay or grow term $e^{-\mathrm{Im}(\epsilon_{n}^{*}-\epsilon_{m})t}$, but this approach neglects the biorthogonal structure of the non-Hermitian Hamiltonian.

In the second manner, we define $\bra{\tilde{\psi}(t)}$ as \cite{yang2024complex}
\begin{align}
    \bra{\tilde{\psi}(t)}=\bra{\psi(0)}e^{i\hat{H}t}=\sum_{n}\tilde{\mathscr{C}}_{n}^{*}(t)\langle\!\langle u_{n}|,
\end{align}
where $\tilde{\mathscr{C}}_{n}^{*}(t)=\braket{\psi(0)|e^{i\hat{H}t}|u_{n}}=\tilde{\mathscr{C}}_{n}^{*}e^{i\epsilon_{n}t}$ with $\tilde{\mathscr{C}}_{n}^{*}=\braket{\psi(0)|u_{n}}$. The Heisenberg equation then retains the same form as in Hermitian systems $\partial_{t}\hat{O}(t)=-i\left[\hat{O}(t),\hat{H}\right]$, and the observable of $\hat{O}(t)$ is 
\begin{align}
    \langle \hat{O}(t) \rangle =\braket{\tilde{\psi}(t)|\hat{O}|\psi(t)}=\sum_{nm}\tilde{\mathscr{C}}_{n}^{*}(t)\mathscr{C}_{m}(t)\langle\!\langle u_{n}|\hat{O}\ket{u_{m}}=\sum_{nm}\tilde{\mathscr{C}}_{n}^{*}\mathscr{C}_{m}e^{i(\epsilon_{n}-\epsilon_{m})t}\langle\!\langle u_{n}|\hat{O}\ket{u_{m}}.
\end{align}
This also requires normalization by  $1/\braket{\tilde{\psi}(t)|\psi(t)}$ at each time due to the decay or grow term $e^{-\mathrm{Im}(\epsilon_{n}-\epsilon_{m})t}$. However, if $\hat{O}$ is diagonal in the biorthogonal eigenstate basis, such as the Hamiltonian itself, the expectation value simplifies to $ \langle \hat{O}(t) \rangle=\sum_{n}\tilde{\mathscr{C}}_{n}^{*}\mathscr{C}_{n}\tilde{O}_{nn}$, which is time-independent and requires normalization by $1/\sum_{n}\tilde{\mathscr{C}}_{n}^{*}\mathscr{C}_{n}$, where $\tilde{O}_{nn}=\langle\!\langle u_{n}|\hat{O}\ket{u_{n}}$, e.g., the eigenenergy $\epsilon_{n}$ for Hamiltonian.

The third manner involves using the complete biorthogonal form, as defined in Ref. \cite{jing2024dyn}, 
\begin{align}
    \ket{\tilde{\psi}(t)}&=\sum_{n}\mathscr{C}_{n}(t)|u_{n}\rangle\!\rangle,\\
    \bra{\tilde{\psi}(t)}&=\sum_{n}\mathscr{C}_{n}^{*}(t)\langle\!\langle u_{n}|.
\end{align}
The expectation value of $\hat{O}(t)$ in this case is
\begin{align}
    \langle \hat{O}(t) \rangle =\braket{\tilde{\psi}(t)|\hat{O}|\psi(t)}=\sum_{nm}\mathscr{C}_{n}^{*}(t)\mathscr{C}_{m}(t)\langle\!\langle u_{n}|\hat{O}\ket{u_{m}}=\sum_{nm}\mathscr{C}_{n}^{*}\mathscr{C}_{m}e^{i(\epsilon_{n}^{*}-\epsilon_{m})t}\langle\!\langle u_{n}|\hat{O}\ket{u_{m}}.
\end{align}
This still requires normalization by $1/\braket{\tilde{\psi}(t)|\psi(t)}$ at each time due to the decay or grow term $e^{-\mathrm{Im}(\epsilon_{n}^{*}-\epsilon_{m})t}$, and we cannot obtain a simple Heisenberg equation from this approach.

\section{Loschmidt amplitudes in non-Hermitian systems}

The origins of non-Hermitian Hamiltonians are diverse, encompassing phenomena such as non-equilibrium open quantum systems \cite{song2019damping}. Here, we set aside these intricate physical contexts and focus directly on the non-Hermitian Hamiltonian itself. By examining the quantum statistical operator (density matrix), we investigate the general form of the non-Hermitian Loschmidt amplitudes and the associated dynamical quantum phase transitions. It is important to note that the non-Hermitian statistical mechanics we consider differs from related works, which primarily study systems where non-Hermitian Hamiltonians are coupled to an external environment \cite{du2022pt, cao2023statistical, cao2024topo}. In contrast, our work focuses on the intrinsic statistical mechanical properties of non-Hermitian Hamiltonians themselves.

Considering the biorthogonal basis nature of non-Hermitian Hamiltonians, we can define the non-Hermitian density matrix upon non-biorthogonal and biorthogonal formulations as 
\begin{align}
    &\rho=\sum_{\alpha}\ket{\Psi_{\alpha}}p_{\alpha}\bra{\Psi_{\alpha}}, \label{suppeqnbdm}\\
    &\varrho=\sum_{\alpha}\ket{\Psi_{\alpha}}p_{\alpha}\langle\!\langle\Psi_{\alpha}| \label{suppeqbdm},
\end{align}
where $\ket{\Psi_{\alpha}}, |\Psi_{\alpha}\rangle\!\rangle$ represent the generic many-body right and left eigenstates, respectively, and $p_{\alpha}$ denote the participation probabilities of these eigenstates. It is important to note that $p_{\alpha}$ can take complex values, which implies that the non-negative trace-class property of the density matrices cannot be guaranteed in non-Hermitian systems. Although complex participation probabilities lack direct physical significance, the inherent ambiguity in the physical interpretation of non-Hermitian Hamiltonians compels us to focus on derived quantities obtained through further analysis to quantify their physical relevance.
Notably, we do not need to consider the case of swapping left and right eigenstates in Eq. (\ref{suppeqbdm}), as it pertains to the study of $\hat{H}^{\dagger}$. Another point that needs to be clarified is that the calculation of the trace must use the biorthogonal basis, even if the density matrix is expressed in a non-biorthogonal form. The trace property in the biorthogonal basis, which for an arbitrary operator $\hat{O}$ takes the form
\begin{align}
    \label{supeqtrace}
    \mathrm{Tr}\left[\hat{O}\right]=\sum_{\alpha}\langle\!\langle\Psi_{\alpha}|\hat{O}\ket{\Psi_{\alpha}}=\sum_{\alpha}\bra{\Psi_{\alpha}}\hat{O}|\Psi_{\alpha}\rangle\!\rangle.
\end{align}
This formulation ensures the consistency of the trace operation in the biorthogonal framework, generalizing the conventional Hermitian case. The last two equalities in Eq. (\ref{supeqtrace}) hold due to the cyclic property of the trace and the biorthogonal structure of the eigenstates, that is $\mathrm{Tr}\left[S^{-1}\hat{O}S\right]=\mathrm{Tr}\left[S^{\dagger}\hat{O}(S^{\dagger})^{-1}\right]$, where the transformation matrix $S$ and its inverse $S^{-1}$ are constructed from the right and left eigenstates of the non-Hermitian Hamiltonian, respectively,
\begin{align}
    S=\left(\begin{matrix}
        \ket{\Psi_{1}}_{1} & \ldots & \ket{\Psi_{N}}_{1}\\
        \vdots & \ddots & \vdots\\
        \ket{\Psi_{1}}_{N} & \ldots & \ket{\Psi_{N}}_{N}
    \end{matrix}
    \right), \quad
    S^{-1}=\left(\begin{matrix}
        \langle\!\langle\Psi_{1}|_{1} & \ldots & \langle\!\langle\Psi_{1}|_{N}\\
        \vdots & \ddots & \vdots\\
        \langle\!\langle\Psi_{N}|_{1} & \ldots & \langle\!\langle\Psi_{N}|_{N}
    \end{matrix}
    \right).
\end{align}

\subsection{Biorthogonal formulation}

Biorthogonalization serves as the natural generalization of the orthogonal framework in Hermitian systems to the non-Hermitian regime, providing a complete basis for describing the eigenstates and dynamics of non-Hermitian Hamiltonians. The base space $\mathcal{H}_{\rho}$ and structure group $\mathbb{G}$ of the principal bundle in biorthogonal formulation are both limitlessly generalized to $\mathrm{GL}(\mathbb{C},N)$, which leads to the total space $\mathcal{H}_{\omega}=\mathrm{GL}(\mathbb{C},N) \times \mathrm{GL}(\mathbb{C},N)$. The surjection is constructed by 
\begin{align}
    p(\omega)=\omega \tilde{\omega},
\end{align}
where 
\begin{align}
    &\omega=\sum_{\alpha}\sqrt{p_{\alpha}}\ket{\Psi_{\alpha}}\bra{\Psi_{\alpha}}U, \label{eq1} \\
    &\tilde{\omega}=U^{-1}\sum_{\alpha}\sqrt{p_{\alpha}}|\Psi_{\alpha}\rangle\!\rangle\langle\!\langle\Psi_{\alpha}|, \label{eq2} \\
    & U\in \mathrm{GL}(\mathbb{C},N) \nonumber . 
\end{align}
The corresponding purification is 
\begin{align}
    \ket{\omega}&=\sum_{\alpha}\sqrt{p_{\alpha}}\ket{\Psi_{\alpha}}\otimes U^{T} \ket{\Psi_{\alpha}}, \\
    \langle\!\langle\tilde{\omega}|&=\sum_{\alpha}\sqrt{p_{\alpha}}\langle\!\langle\Psi_{\alpha}| \otimes \langle\!\langle\Psi_{\alpha}| (U^{T})^{-1}.
\end{align}
The associated calculation are given by
\begin{align}
    \varrho&=\mathrm{Tr}_{2}\left[\ket{\omega}\langle\!\langle\tilde{\omega}|\right]=\sum_{\alpha, \beta, \gamma}\sqrt{p_{\alpha}}\sqrt{p_{\beta}}\ket{\Psi_{\alpha}}\otimes \langle\!\langle\Psi_{\gamma}|U^{T} \ket{\Psi_{\alpha}}\langle\!\langle\Psi_{\beta}| \otimes \langle\!\langle\Psi_{\beta}| (U^{T})^{-1} \ket{\Psi_{\gamma}} \nonumber\\
    &=\sum_{\alpha, \beta, \gamma}\sqrt{p_{\alpha}}\sqrt{p_{\beta}}\ket{\Psi_{\alpha}}\langle\!\langle\Psi_{\beta}|\otimes \langle\!\langle\Psi_{\gamma}|U^{T} \ket{\Psi_{\alpha}} \langle\!\langle\Psi_{\beta}| (U^{T})^{-1} \ket{\Psi_{\gamma}} =\sum_{\alpha, \beta, \gamma}\sqrt{p_{\alpha}}\sqrt{p_{\beta}}\ket{\Psi_{\alpha}}\langle\!\langle\Psi_{\beta}|\otimes \langle\!\langle\Psi_{\beta}| (U^{T})^{-1} \ket{\Psi_{\gamma}}\langle\!\langle\Psi_{\gamma}|U^{T} \ket{\Psi_{\alpha}} \nonumber\\
    &=\sum_{\alpha, \beta}\sqrt{p_{\alpha}}\sqrt{p_{\beta}}\ket{\Psi_{\alpha}}\langle\!\langle\Psi_{\beta}|\otimes \langle\!\langle\Psi_{\beta}| (U^{T})^{-1} U^{T} \ket{\Psi_{\alpha}} =\sum_{\alpha}\ket{\Psi_{\alpha}}p_{\alpha}\langle\!\langle\Psi_{\alpha}|,
\end{align}
\begin{align}
    \mathrm{Tr}\left[\tilde{\omega}\omega'\right]&=\mathrm{Tr}\left[\sum_{\alpha,\beta}\sqrt{p_{\alpha}}\sqrt{p_{\beta}'}U^{-1}|\Psi_{\alpha}\rangle\!\rangle\langle\!\langle\Psi_{\alpha}\ket{\Psi_{\beta}}\bra{\Psi_{\beta}}U\right]\nonumber\\
    &=\mathrm{Tr}\left[\sum_{\alpha,\beta}\sqrt{p_{\alpha}}\sqrt{p_{\beta}'}|\Psi_{\alpha}\rangle\!\rangle\langle\!\langle\Psi_{\alpha}\ket{\Psi_{\beta}}\bra{\Psi_{\beta}}\right]\nonumber\\
    &=\mathrm{Tr}\left[\sum_{\alpha}\sqrt{p_{\alpha}}\sqrt{p_{\alpha}'}|\Psi_{\alpha}\rangle\!\rangle\bra{\Psi_{\alpha}}\right] \nonumber\\
    &=\sum_{\alpha,\beta}\sqrt{p_{\alpha}}\sqrt{p_{\alpha}'}\bra{\Psi_{\beta}}\Psi_{\alpha}\rangle\!\rangle\bra{\Psi_{\alpha}}\Psi_{\beta}\rangle\!\rangle \nonumber\\
    &=\sum_{\alpha}\sqrt{p_{\alpha}}\sqrt{p_{\alpha}'},
\end{align}
\begin{align}
    \mathrm{Tr}\left[\omega'\tilde{\omega}\right]&=\mathrm{Tr}\left[\sum_{\alpha,\beta}\sqrt{p_{\alpha}'}\sqrt{p_{\beta}}\ket{\Psi_{\alpha}}\bra{\Psi_{\alpha}}UU^{-1}|\Psi_{\beta}\rangle\!\rangle\langle\!\langle\Psi_{\beta}|\right]\nonumber\\
    &=\mathrm{Tr}\left[\sum_{\alpha,\beta}\sqrt{p_{\alpha}'}\sqrt{p_{\beta}}\ket{\Psi_{\alpha}}\bra{\Psi_{\alpha}}\Psi_{\beta}\rangle\!\rangle\langle\!\langle\Psi_{\beta}|\right]\nonumber\\
    &=\mathrm{Tr}\left[\sum_{\alpha}\sqrt{p_{\alpha}'}\sqrt{p_{\alpha}}\ket{\Psi_{\alpha}}\langle\!\langle\Psi_{\alpha}|\right] \nonumber\\
        &= \sum_{\alpha,\beta}\sqrt{p_{\alpha}'}\sqrt{p_{\alpha}}\langle\!\langle\Psi_{\beta}\ket{\Psi_{\alpha}}\langle\!\langle\Psi_{\alpha}\ket{\Psi_{\beta}}\nonumber\\
        &=\sum_{\alpha}\sqrt{p_{\alpha}'}\sqrt{p_{\alpha}},
\end{align}
\begin{align}
    \langle\!\langle\tilde{\omega}\ket{\omega'}&=\sum_{\alpha,\beta}\sqrt{p_{\alpha}}\sqrt{p_{\beta}'}\langle\!\langle\Psi_{\alpha}| \otimes \langle\!\langle\Psi_{\alpha}| (U^{T})^{-1}\ket{\Psi_{\beta}}\otimes U^{T} \ket{\Psi_{\beta}}\nonumber\\
    &=\sum_{\alpha,\beta}\sqrt{p_{\alpha}}\sqrt{p_{\beta}'}\langle\!\langle\Psi_{\alpha}\ket{\Psi_{\beta}}\otimes  \langle\!\langle\Psi_{\alpha}| (U^{T})^{-1} U^{T} \ket{\Psi_{\beta}}\nonumber\\
    &=\sum_{\alpha}\sqrt{p_{\alpha}}\sqrt{p_{\alpha}'},
\end{align}
namely, $\langle\!\langle\tilde{\omega}\ket{\omega'}=\mathrm{Tr}\left[\omega'\tilde{\omega}\right]=\mathrm{Tr}\left[\tilde{\omega}\omega'\right]$. 

However, due to the mathematical properties of complex-valued quantities, defining parallel transport in non-Hermitian systems in a way that minimizes the distance is challenging, as constructing a positive-definite distance analogous to Eq. (\ref{suppeqdis}) is difficult. Furthermore, such a definition cannot be directly derived from the in-phase condition commonly used in quantum interferometry experiments \cite{ajoqvist2000}. In fact, the precise form of the parallel transport condition does not need to be explicitly defined. We can assume the existence of parallel transport paths analogous to those in Hermitian systems. As in Hermitian regime, the parallel transport evolution $V(t)$ is uniquely generated by the connection $\mathscr{A}$ on the principal bundles \cite{nakahara2003book}
\begin{align}
    V(t)=\mathcal{P}e^{\int_{0}^{t} dt' \mathscr{A}(t')},
\end{align}
which results in an group element of $\mathrm{GL}(\mathbb{C},N)$ acting on the fibre and $\omega(t)=\omega(0)V(t)$. If we define $s(t)\equiv \sum_{\alpha}\sqrt{p_{\alpha}}\ket{\Psi_{\alpha}}\bra{\Psi_{\alpha}}$ as a section from $\mathcal{H}_{\varrho}$ to $\mathcal{H}_{\omega}$, this implies that the initial gauge is given by $\omega(0) = s(t) U_{g}(0), \tilde{\omega}(0)=U_{g}(0)^{-1}\tilde{s}(t)$, and $\tilde{s}(t)=\sum_{\alpha}\sqrt{p_{\alpha}}|\Psi_{\alpha}\rangle\!\rangle\langle\!\langle\Psi_{\alpha}|$ where $U_{g}(0) = 1$. Consequently, the parallel transport specifies $\omega(t) = s(t) V(t)$. As analog to Hermitian systems in Fig. \ref{supfig-gauge}, any initial gauge choice $U_{g}(0)$ specifies a section $s(t)U_{g}(t)$ and leads to $\omega(t)=s(t)V(t)U_{g}(0)$, that is $\omega(t)=s(t)U_{g}(t)V_{g}(t)$, where $V_{g}(t)=U_{g}^{-1}(t)V(t)U_{g}(0)$. The holonomy group, the subgroup of $\mathrm{GL}(\mathbb{C},N)$, is constituted by the collection of $V(t)$ along some closed parallel transport paths \cite{nakahara2003book}. Even though the path is not closed, the phase 
\begin{align}
\label{supeqgeo}
    \phi_{g}=\arg \mathrm{Tr}\left[\tilde{\omega}(0)\omega(t)\right],
\end{align}
remains gauge invariant. This phase represents a purely $\mathrm{GL}(\mathbb{C},N)$-invariant geometric phase along any parallel transport path and corresponds to the non-Hermitian Uhlmann phase in the case of a closed path.  

Analog to Eq. (\ref{suppeqlsa}), we can define the biorthogonal Loschmidt amplitude for mixed states in non-Hermitian systems as 
\begin{align}
    \label{suppeqnhlsa}
    \mathcal{G}(t)=\mathrm{Tr}\left[\varrho(0)U(t)\right],
\end{align}
where $\varrho(0)$ represents the density matrix of the prequench Hamiltonian $\hat{H}_{0}$, and $U(t)$ is the time evolution operator of the postquench Hamiltonian $\hat{H}_{1}$. According to Eqs. (\ref{eq1}) and (\ref{eq2}), the time evolution of $\omega,\tilde{\omega}$ is given by
\begin{align}
    \omega(t)&=U(t)\omega(0)U_{g}, \\
    \tilde{\omega}(t)&=U_{g}^{-1}\tilde{\omega}(0)U(t)^{-1}.
\end{align}
Assuming the identity initial gauge $\varrho(0)=s(0)\tilde{s}(0)$, the biorthogonal mixed-state Loschmidt amplitude can be written 
\begin{align}
    \mathcal{G}(t)&=\mathrm{Tr}\left[s(0)\tilde{s}(0)U(t)\right]\nonumber\\
    &=\mathrm{Tr}\left[U(t)^{-1}s(t)(U(t)^{\dagger})^{-1}\tilde{s}(0)U(t)\right]\nonumber\\
     &=\mathrm{Tr}\left[\tilde{s}(0)s(t)(U(t)^{\dagger})^{-1}\right],
\end{align}
which implies that $\arg \mathcal{G}(t)$ is $\mathrm{GL}(\mathbb{C},N)$-gauge invariant if $(U(t)^{\dagger})^{-1}=V(t)$ corresponding the parallel transport. For generic time evolution, we have
\begin{align}
    \langle\!\langle\tilde{\omega}(0)\ket{\omega(t)}&=\sum_{\alpha,\beta}\sqrt{p_{\alpha}}\sqrt{p_{\beta}}\langle\!\langle\Psi_{\alpha}| \otimes \langle\!\langle\Psi_{\alpha}| (U^{T})^{-1}U(t)\ket{\Psi_{\beta}}\otimes U^{T} \ket{\Psi_{\beta}}  \nonumber\\
    &=\sum_{\alpha,\beta}\sqrt{p_{\alpha}}\sqrt{p_{\beta}}\langle\!\langle\Psi_{\alpha}|U(t)\ket{\Psi_{\beta}}\otimes \langle\!\langle\Psi_{\alpha}| (U^{T})^{-1} U^{T} \ket{\Psi_{\beta}}  \nonumber\\
    &=\sum_{\alpha}p_{\alpha}\langle\!\langle\Psi_{\alpha}|U(t)\ket{\Psi_{\alpha}}  \nonumber\\
    &=\mathrm{Tr}\left[\varrho(0)U(t)\right],
\end{align}
leading to $\mathrm{Tr}\left[\varrho(0)U(t)\right]=\mathrm{Tr}\left[\tilde{\omega}(0)\omega(t)\right]$. If we impose $\mathbb{U}(1)$ gauge transformation $\ket{(\omega(t))}\rightarrow \ket{(\omega'(t))}=e^{i\vartheta(t)}\ket{(\omega(t))}, \langle\!\langle\tilde{\omega}(0)|\rightarrow \langle\!\langle\tilde{\omega}'(0)|=\langle\!\langle\tilde{\omega}(0)| e^{-i \vartheta(0)}$, 
\begin{align}
    \arg \langle\!\langle\tilde{\omega}'(0)\ket{\omega'(t)}=\arg \langle\!\langle\tilde{\omega}(0)\ket{\omega(t)}+[\vartheta(t)-\vartheta(0)].
\end{align}
To compensating this change, we should introduce the dynamical phase as in Hermitian systems, 
\begin{align}
    \phi_{dyn}=-\int_{0}^{t}\langle\!\langle\tilde{\omega}(t)|\hat{H}\ket{\omega(t)},
\end{align}
such that
\begin{align}
    \int_{0}^{t}\langle\!\langle\tilde{\omega}'(t)|\hat{H}\ket{\omega'(t)}=
    \int_{0}^{t}\langle\!\langle\tilde{\omega}(t)|\hat{H}\ket{\omega(t)}-[\vartheta(t)-\vartheta(0)].
\end{align}
Hence, we obtain the $\mathbb{U}(1)$-invariant geometric phase
\begin{align}
    \phi=\arg \mathcal{G}(t)-\phi_{dyn}.
\end{align}

In fact, the non-Hermitian Loschmidt amplitude in Eq. (\ref{suppeqnhlsa}) simply constitutes a biorthogonal generalization of its Hermitian counterpart in Eq. (\ref{suppeqlsa}). However, the connection between parallel transport and the Loschmidt amplitude is not mandatory. Moreover, the non-Hermitian extension of geometric phases generally lacks a well-defined quantization scheme in principle, thus failing to serve as a reliable indicator for DQPTs. Although certain numerical studies have demonstrated the dynamical order-parameter characteristics of non-Hermitian geometric phases \cite{zhou2018dyn, zhou2021non, mondal2022dyn, mondal2023dyn, jing2024dyn}, this work will focus exclusively on the fundamental theoretical relationship between Loschmidt amplitudes and DQPTs.

{\bf{\em Pure-state Loschmidt amplitude.---}}
Before proceeding with the biorthogonal mixed-state Loschmidt amplitude in non-Hermitian systems, we first clarify its pure-state counterpart. The initial state of $\hat{H}_{0}$ is prepared by 
\begin{align}
    \varrho(0)=\ket{\Psi_{0}}\langle\!\langle \Psi_{0}|,
\end{align}
where $\ket{\Psi_{0}}$ ($|\Psi_{0}\rangle\!\rangle$) is the right (left) ground state. Assuming that we are studying a one-dimensional (1D) non-interacting momentum-space Hamiltonian $H_{0}(k)$ with well-defined occupied bands, the right and left ground states are expressed by 
\begin{align}
    \ket{\Psi_{0}}&=\bigotimes_{(n,k)\in occ} \ket{u_{n}^{0}(k)}, \label{focksr}\\
    |\Psi_{0}\rangle\!\rangle&= \bigotimes_{(n,k)\in occ} |u_{n}^{0}(k)\rangle\!\rangle, \label{focksl}
\end{align}
where $n$ labels the band index. If we adapt the second manner of non-Hermitian dynamic described above, we arrive at $\varrho(t)=U(t)^{-1}\varrho(0)U(t)$ after the quantum quench by $\hat{H}_{1}$, where $U(t)=e^{-i\hat{H}_{1}t}$. Consequently, 
\begin{align}
    \ket{u_{n}^{0}(k,t)} &=e^{-iH_{1}(k)t}\ket{u_{n}^{0}(k)},\\
    |u_{n}^{0}(k,t)\rangle\!\rangle &=e^{-iH_{1}^{\dagger}(k)t}|u_{n}^{0}(k)\rangle\!\rangle\equiv \ket{\tilde{u}_{n}^{0}(k,t)},\\
    \langle\!\langle u_{n}^{0}(k,t)|&=\langle\!\langle u_{n}^{0}(k)|e^{iH_{1}(k)t}\equiv \bra{\tilde{u}_{n}^{0}(k,t)}.
\end{align}
The biorthogonal Loschmidt amplitude is calculated by
\begin{align}
    \mathcal{G}(t)&=\mathrm{Tr}\left[\ket{\Psi_{0}}\langle\!\langle \Psi_{0}|e^{-i\hat{H}_{1}t}\right] \nonumber\\
    &=\prod_{(n,k) \in occ}\prod_{(n',k') \in occ}\prod_{k_{0}\in \mathrm{BZ}}\langle\!\langle u_{n'}^{0}(k') \ket{u_{n}^{0}(k)}\langle\!\langle u_{n}^{0}(k)|e^{-i \hat{c}_{k_{0}}^{\dagger}H_{1}(k_{0})\hat{c}_{k_{0}}t}\ket{u_{n'}^{0}(k')}\nonumber\\
    &=\prod_{(n,k)\in occ}\langle\!\langle u_{n}^{0}(k)|e^{-i H_{1}(k)t}\ket{u_{n}^{0}(k)}\nonumber\\
    &=\prod_{(n,k)\in occ}\langle\!\langle u_{n}^{0}(k)\ket{u_{n}^{0}(k,t)}.
\end{align}
The Loschmidt echo is defined by 
\begin{align}
    \mathcal{L}(t)=\left|\mathcal{G}(t)\right|^{2}=\mathcal{G}(t)\mathcal{G}(t)^{*}.
\end{align}
Note that this Loschmidt echo is consistent with that given in Ref. \cite{jing2024dyn}, which is based the third manner of non-Hermitian dynamic described above. In Ref. \cite{jing2024dyn}, the associated state of 
\begin{align}
    \ket{u_{n}^{0}(k,t)}=e^{-iH_{1}(k)t}\ket{u_{n}^{0}(k)}=\sum_{m}\ket{u_{m}^{0}(k)}\langle\!\langle u_{m}^{0}(k) \ket{u_{n}^{0}(k,t)}=\sum_{m}\mathscr{C}_{mn}(k,t)\ket{u_{m}^{0}(k)},
\end{align}
is defined by 
\begin{align}
    |u_{n}^{0}(k,t)\rangle\!\rangle&=\sum_{m}|u_{m}^{0}(k)\rangle\!\rangle \langle\!\langle u_{m}^{0}(k)\ket{u_{n}^{0}(k,t)}=\sum_{m}\mathscr{C}_{mn}(k,t)|u_{m}^{0}(k)\rangle\!\rangle\equiv \ket{\tilde{u}_{n}^{0}(k,t)},\nonumber\\
    \langle\!\langle u_{n}^{0}(k,t)|&=\sum_{m}\langle\!\langle u_{m}^{0}(k)|\mathscr{C}_{mn}^{*}(k,t)\equiv  \langle\!\langle \tilde{u}_{n}^{0}(k,t)|.
\end{align}
In this framework, we remark that
\begin{align}
    \langle\!\langle u_{n}^{0}(k)\ket{u_{n}^{0}(k,t)}&=\langle\!\langle u_{n}^{0}(k)\sum_{m}\mathscr{C}_{mn}(k,t)\ket{u_{m}^{0}(k)}=\mathscr{C}_{nn}(k,t),\nonumber\\
    \langle\!\langle u_{n}^{0}(k,t)\ket{u_{n}^{0}(k)}&=\sum_{m}\langle\!\langle u_{m}^{0}(k)|\mathscr{C}_{mn}^{*}(k,t)\ket{u_{n}^{0}(k)}=\mathscr{C}^{*}_{nn}(k,t),
\end{align}
giving 
\begin{align}
    \langle\!\langle u_{n}^{0}(k)\ket{u_{n}^{0}(k,t)}=\left(\langle\!\langle u_{n}^{0}(k,t)\ket{u_{n}^{0}(k)}\right)^{*}.
\end{align}
Hence, the Loschmidt echo is given by
\begin{align}
    \mathcal{L}(t)=\prod_{(n,k)\in occ}\langle\!\langle u_{n}^{0}(k)\ket{u_{n}^{0}(k,t)}\langle\!\langle u_{n}^{0}(k,t)\ket{u_{n}^{0}(k)},
\end{align}
and after normalizing by the denominator $\langle\!\langle u_{n}^{0}(k,t)\ket{u_{n}^{0}(k,t}$, we obtain
\begin{align}
    \label{suppeqechobi}
    \mathcal{L}(t)=\prod_{(n,k)\in occ}\frac{\langle\!\langle u_{n}^{0}(k)\ket{u_{n}^{0}(k,t)}\langle\!\langle u_{n}^{0}(k,t)\ket{u_{n}^{0}(k)}}{\langle\!\langle u_{n}^{0}(k,t)\ket{u_{n}^{0}(k,t)}},
\end{align}
which precisely corresponds to the form of the Loschmidt echo for the two-band case with the lower occupied band, as presented in Ref. \cite{jing2024dyn}.
We emphasize that the results of the Loschmidt amplitude and echo are independent of the choice between the second and third manners of non-Hermitian dynamics, which are up to normalization factors at each time. The advantage of the third manner, along with the biorthogonal formulation in Ref. \cite{jing2024dyn}, lies in its normalized and symmetric form of the Loschmidt echo, as presented in Eq. (\ref{suppeqechobi}), which can arrive at the form 
\begin{align}
    \mathcal{L}(t)=\prod_{(n,k)\in occ}\frac{\left|\mathscr{C}_{nn}(k,t)\right|^{2}}{\sum_{m}\left|\mathscr{C}_{mn}(k,t)\right|^{2}}.
\end{align}

{\bf{\em Mixed-state Loschmidt amplitude (Fock space formulation).---}} 
A right and left many-body state in Fock space are expressed as Eqs. (\ref{focksr}) and (\ref{focksl}), respectively. The trace of an operator $\hat{O}$ in the biorthogonal basis is given by 
\begin{align}
    \mathrm{Tr}\left[\hat{O}\right]&=\sum_{\alpha}\langle\!\langle \Psi_{\alpha}|\hat{O}\ket{\Psi_{\alpha}}\nonumber\\
    &=\sum_{\left\{\tilde{k},\tilde{n}\right\}}\bigotimes_{k,n}\langle\!\langle u_{n}(k)|\hat{O}\ket{u_{n}(k)},
\end{align}
where $\sum_{\left\{\tilde{k},\tilde{n}\right\}}=\sum_{\alpha}$ denotes the summation over all possible Fock states (labeled by $\alpha$) with the corresponding occupied momenta $k$ and band indices $n$. The initial density matrix can be decomposed as
\begin{align}
   \label{suppeqbdensity}
    \varrho(0)&=\sum_{\alpha}p_{\alpha}\left(\bigotimes_{k,n}\ket{u_{n}^{0}(k)}\right)\left(\bigotimes_{k,n}\langle\!\langle u_{n}^{0}(k)|\right) \nonumber\\
    &=\sum_{\left\{\tilde{k},\tilde{n}\right\}}\bigotimes_{j}p_{k_j}^{n_j}\ket{u_{n_j}^{0}(k_j)}\langle\!\langle u_{n_j}^{0}(k_j)|\nonumber\\
    &=\sum_{\left\{\tilde{k}\right\}}\sum_{\left\{\tilde{n}\right\}}\bigotimes_{j}p_{k_j}^{n_j}\ket{u_{n_j}^{0}(k_j)}\langle\!\langle u_{n_j}^{0}(k_j)|\nonumber\\
    &=\sum_{\left\{\tilde{k}\right\}}\bigotimes_{j}\sum_{n}p_{k_j}^{n}\ket{u_{n}^{0}(k_j)}\langle\!\langle u_{n}^{0}(k_j)|\nonumber\\
     &=\sum_{\left\{\tilde{k}\right\}}\bigotimes_{j}\varrho_{k_j}(0),
\end{align}
where $p_\alpha=\prod_{j}p_{k_j}^{n_j}$ and $\varrho_{k_j}(0)\equiv\sum_{n}p_{k_j}^{n}\ket{u_{n}^{0}(k_j)}\langle\!\langle u_{n_j}^{0}(k_j)|$. Remarkably, the summation or product over the indices $n$ and $k$ in Eq. (\ref{suppeqbdensity}) is, in principle, arbitrary and may include repeated entries. In fermionic systems, such repetitions lead to vanishing contributions due to the Pauli exclusion principle, while in both bosonic and fermionic cases, proper normalization requires symmetrization or anti-symmetrization of the many-body states \cite{negele2018book}. Hence, Eq. (\ref{suppeqbdensity}) should be understood as a formal expression. In this work, we focus on a simple fermionic setting in which a specific, non-repeating momentum occupation configuration---such as half or the entire Brillouin zone (BZ)---is predetermined. Consequently, the summation over all possible momentum configurations and antisymmetrization is omitted in Eq. (\ref{suppeqbdensity}). The approach for the non-biorthogonal formulation proceeds in a similar fashion thereafter.

Assuming that all band indices $n$ are occupied for fermionic systems, we simplify 
\begin{align}
    \varrho(0)&=\bigotimes_{k\in \mathrm{BZ}/2}\varrho_{k}(0), \quad \text{half-filled}, \\
    \varrho(0)&=\bigotimes_{k\in \mathrm{BZ}}\varrho_{k}(0), \quad \text{full-filled},
\end{align}
where we have specified a particular non-repeating momentum occupation. Consequently, the Loschmidt amplitude for the particular BZ occupation is given by (up to normalization coefficient)
\begin{align}
    \mathcal{G}(t)&=\mathrm{Tr}\left[\varrho(0)U(t)\right]\nonumber\\
    &=\mathrm{Tr}\left[\bigotimes_{k}\varrho_{k}(0)\prod_{k_0}e^{-i \hat{c}_{k_{0}}^{\dagger}H_{1}(k_{0})\hat{c}_{k_{0}}t}\right]\nonumber\\
    &=\prod_{k}\mathcal{G}_{k}(t),
\end{align}
where 
\begin{align}
    \mathcal{G}_{k}(t)&=\mathrm{Tr}\left[\varrho_{k}(0)U_{k}(t)\right]\nonumber\\
     &=\sum_{k',n'}\langle\!\langle u_{n'}(k')|\left[\sum_{n}p_{k}^{n}\ket{u_{n}^{0}(k)}\langle\!\langle u_{n}^{0}(k)|e^{-i \hat{c}_{k_{0}}^{\dagger}H_{1}(k_{0})\hat{c}_{k_{0}}t}\right]\ket{u_{n'}(k')}\nonumber\\
    &=\sum_{n}p_{k}^{n}\langle\!\langle u_{n}^{0}(k)|e^{-i H_{1}(k)t} \ket{u_{n}^{0}(k)}\nonumber\\
    &=\sum_{n}p_{k}^{n}\langle\!\langle u_{n}^{0}(k) \ket{u_{n}^{0}(k,t)}, \\
    U_{k}(t)&=e^{-i H_{1}(k)t}.
\end{align}
It can be seen that this formula is the biorthogonal generalization of the mixed-state case for Hermitian systems, as presented in Ref. \cite{bhattacharya2017mixed}. The corresponding mixed-state Loschmidt echo is given by
\begin{align}
    \mathcal{L}(t)=\left|\mathcal{G}(t)\right|^{2}=\prod_{k}\sum_{n,n'}p_{k}^{n*}p_{k}^{n'}\langle\!\langle u_{n}^{0}(k,t)\ket{u_{n}^{0}(k)}\langle\!\langle u_{n'}^{0}(k)\ket{u_{n'}^{0}(k,t)}.
\end{align}  
Since the Fisher zeros that signal DQPTs are invariant under normalization in the denominator, the overall normalization factor of the Loschmidt amplitude or echo can be safely omitted. 

For example of a two-band model $\mathcal{H}_{0}(k)=\vec{h}_{0}(k)\cdot \vec{\sigma}$, with $\vec{\sigma}=(\sigma_{x},\sigma_{y},\sigma_{z})$, we specify the initial state as the generalized Gibbs state with the temperature $T=1/\beta$,
\begin{align}
    \varrho_{k}(0)&=\frac{e^{-\beta \mathcal{H}_{0}(k)}}{\mathrm{Tr}\left[e^{-\beta \mathcal{H}_{0}(k)}\right]}\nonumber\\
    &=\frac{\cosh\left[\beta h_{0}(k)\right]\mathbbm{1}-\sinh\left[\beta h_{0}(k)\right]\hat{h}_{0}(k)\cdot \vec{\sigma}}{2\cosh\left[\beta h_{0}(k)\right]}\nonumber\\
    &=\frac{1}{2}\left[\mathbbm{1}-n_{k}\hat{h}_{0}(k)\cdot \vec{\sigma}\right],
\end{align}
where $h_{0}(k)=\left(h_{0,x}^{2}(k)+h_{0,y}^{2}(k)+h_{0,z}^{2}(k)\right)^{1/2}$, $\hat{h}_{0}(k)=\vec{h}_{0}(k)/h_{0}(k)$, and $n_{k}=\tanh\left[\beta h_{0}(k)\right]$. In addition, it is explicit that 
\begin{align}
    p_{k}^{\pm}&=\frac{e^{-\beta \epsilon_{\pm}^{0}(k)}}{e^{-\beta h_{0}(k)}+e^{\beta h_{0}(k)}}=\frac{1}{2}\left(1\mp n_{k} \right),
\end{align}
where the eigenenergies of two bands are $\epsilon_{k\pm}^{0}=\pm h_{0}(k)=\pm\epsilon_{k}^{0}$. 

In detail, the non-Hermitian two-band model generally reads
\begin{align}
    \mathcal{H}(k)&=h_x(k)\sigma_x+h_y(k) \sigma_y+h_z(k) \sigma_z \nonumber\\
    &= \left(\begin{matrix}
       h_{z} & h_x-i h_y \\
        h_x + ih_y & -h_{z}
    \end{matrix}\right)\nonumber\\
    &=\left(\begin{matrix}
       h_{zr}+ih_{zi} & h_{xr}+ih_{xi}-i(h_{yr}+ih_{yi}) \\
       h_{xr}+ih_{xi}+i(h_{yr}+ih_{yi}) & -(h_{zr}+ih_{zi})
    \end{matrix}\right)\nonumber\\
    &=\left(\begin{matrix}
       h_{zr}+ih_{zi} & h_{xr}+h_{yi}+i(h_{xi}-h_{yr}) \\
       h_{xr}-h_{yi}+i(h_{xi}+h_{yr}) & -(h_{zr}+ih_{zi})
    \end{matrix}\right)\nonumber\\
    &=\left(\begin{matrix}
       G+iF & A+iB \\
       C+iD & -(G+iF)
    \end{matrix}\right),
\end{align}
where
\begin{align}
    A&=h_{xr}+h_{yi},\\
    B&=h_{xi}-h_{yr},\\
    C&=h_{xr}-h_{yi},\\
    D&=h_{xi}+h_{yr},\\
    G&=h_{zr},\\
    H&=h_{zi}.
\end{align}
The energies and corresponding biorthogonal eigenstates are
\begin{align}
    &\epsilon_{\pm}=\pm\sqrt{(A+iB)(C+iD)+(G+iF)^{2}},\\
    & \epsilon_{\pm}^{*}=\pm\sqrt{(A-iB)(C-iD)+(G-iF)^{2}},\\
    &\ket{u_{\pm}(k)}=\frac{1}{\sqrt{2\epsilon_{\pm}\left[\epsilon_{\pm}-(G+iF)\right]}}\left(\begin{matrix}
       A+iB\\
        \epsilon_{\pm}-(G+iF)
    \end{matrix}\right),\\
    &\langle\!\langle u_{\pm}(k)|=\frac{1}{\sqrt{2 \epsilon_{\pm}\left[\epsilon_{\pm}-(G+iF)\right]}}\left(\begin{matrix}
       C+iD &  \epsilon_{\pm}-(G+iF)
    \end{matrix}\right).
\end{align}
The biorthogonal property is given by
\begin{align}
    \langle\!\langle u_{\pm}(k)\ket{u_{\pm}(k)}&=1,\\
    \langle\!\langle u_{\pm}(k)\ket{u_{\mp}(k)}&=0,\\
     \ket{u_{+}(k)}\langle\!\langle u_{+}(k)|+ \ket{u_{-}(k)}\langle\!\langle u_{-}(k)|&=\mathbbm{1}, \\
    \ket{u_{+}(k)}\langle\!\langle u_{+}(k)|- \ket{u_{-}(k)}\langle\!\langle u_{-}(k)|&=\hat{\mathcal{H}}(k)
\end{align}
    
Quenching from $\mathcal{H}_{0}(k)$ to $\mathcal{H}_{1}(k)=\vec{h}_{1}(k)\cdot \vec{\sigma}$, the single-particle time evolution operator is given by
\begin{align}
    U_{k}(t)&=e^{-i \mathcal{H}_{1}(k)t} = \sum_{n=1}^{+\infty}\frac{(-i)^{n}\left[h_{1}(k)\hat{\mathcal{H}}_{1}(k)t\right]^{n}}{n!} \nonumber\\
    &=\sum_{n=0}^{+\infty}\frac{(-i)^{2n}\left[h_{1}(k)t\right]^{2n}}{(2n)!}(\mathscr{P}_{k+}^{1}-\mathscr{P}_{k-}^{1})^{2n}+\sum_{n=0}^{+\infty}\frac{(-i)^{2n+1}\left[h_{1}(k)t\right]^{2n+1}}{(2n+1)!}(\mathscr{P}_{k+}^{1}-\mathscr{P}_{k-}^{1})^{2n+1} \nonumber\\
    &=\sum_{n=0}^{+\infty}\frac{(-1)^{n}\left[h_{1}(k)t\right]^{2n}}{(2n)!}+\sum_{n=0}^{+\infty}\frac{-i(-1)^{n}\left[h_{1}(k)t\right]^{2n+1}}{(2n+1)!}(\mathscr{P}_{k+}^{1}-\mathscr{P}_{k-}^{1}) \nonumber\\
    &=\cos\left[h_{1}(k)t\right]\mathbbm{1}-i\sin\left[h_{1}(k)t\right]\hat{\mathcal{H}}_{1}(k),
\end{align}
where $ h_{1}(k)=\left(h_{1,x}^{2}(k)+h_{1,y}^{2}(k)+h_{1,z}^{2}(k)\right)^{1/2}=\epsilon_{k}^{1}$, $\hat{\mathcal{H}}_{1}(k)=\mathcal{H}_{1}(k)/h_{1}(k)=\mathscr{P}_{k+}^{1}-\mathscr{P}_{k-}^{1}$, and $\mathscr{P}_{k\pm}^{1}=\ket{u_{\pm}^{1}(k)}\langle\!\langle u_{\pm}^{1}(k)|$ is the projection operator. We can then obtain the relation
\begin{align}
    &\langle\!\langle u_{\pm}^{0}(k)|\hat{\mathcal{H}}_{1}(k) \ket{u_{\pm}^{0}(k)}=\frac{1}{\epsilon_{k}^{1}}\langle\!\langle u_{\pm}^{0}(k)|\mathcal{H}_{1}(k) \ket{u_{\pm}^{0}(k)}=\pm\hat{h}_{0} \cdot \hat{h}_{1}
\end{align}
after some algebra.
Hence, the Loschmidt amplitude is (up to normalization coefficient)
\begin{align}
    \mathcal{G}_{k}(t)&=p_{k}^{+}\langle\!\langle u_{+}^{0}(k)|e^{-i \mathcal{H}_{1}(k)t} \ket{u_{+}^{0}(k)}+p_{k}^{-}\langle\!\langle u_{-}^{0}(k)|e^{-i \mathcal{H}_{1}(k)t} \ket{u_{-}^{0}(k)}\nonumber\\
    &=p_{k}^{+}\langle\!\langle u_{+}^{0}(k)|\cos\left[h_{1}(k)t\right]\mathbbm{1}-i\sin\left[h_{1}(k)t\right]\hat{\mathcal{H}}_{1}(k) \ket{u_{+}^{0}(k)}+p_{k}^{-}\langle\!\langle u_{-}^{0}(k)|\cos\left[h_{1}(k)t\right]\mathbbm{1}-i\sin\left[h_{1}(k)t\right]\hat{\mathcal{H}}_{1}(k) \ket{u_{-}^{0}(k)}\nonumber\\
    &=p_{k}^{+}\left\{\cos\left[h_{1}(k)t\right]-i\sin\left[h_{1}(k)t\right]\langle\!\langle u_{+}^{0}(k)|\hat{\mathcal{H}}_{1}(k) \ket{u_{+}^{0}(k)}\right\}+p_{k}^{-}\left\{\cos\left[h_{1}(k)t\right]-i\sin\left[h_{1}(k)t\right]\langle\!\langle u_{-}^{0}(k)|\hat{\mathcal{H}}_{1}(k) \ket{u_{-}^{0}(k)}\right\}\nonumber\\
    &=p_{k}^{+}\left\{\cos\left[h_{1}(k)t\right]-i\sin\left[h_{1}(k)t\right]\hat{h}_{0} \cdot \hat{h}_{1}\right\}+p_{k}^{-}\left\{\cos\left[h_{1}(k)t\right]+i\sin\left[h_{1}(k)t\right]\hat{h}_{0} \cdot \hat{h}_{1}\right\}\nonumber\\
    &=\cos\left[h_{1}(k)t\right](p_k^++p_k^-)-i\sin\left[h_{1}(k)t\right](p_k^+-p_k^-)\hat{h}_{0} \cdot \hat{h}_{1}\nonumber\\
    &=\cos\left[h_{1}(k)t\right]+i\sin\left[h_{1}(k)t\right]n_{k}\hat{h}_{0} \cdot \hat{h}_{1},
\end{align}
which is exactly the result in Ref. \cite{mondal2023dyn}. In the zero-temperature limit $\beta \rightarrow \infty$ (pure ground state case), $n_{k}=\tanh\left[\beta h_{0}(k)\right] \rightarrow 1$, thus $\mathcal{G}_{k}(t)=\cos\left[h_{1}(k)t\right]+i\sin\left[h_{1}(k)t\right]\hat{h}_{0} \cdot \hat{h}_{1}$.

{\em \textbf{Fisher zeros.---}}
The zeros of $\mathcal{G}(t)$ corresponds to the critical points of DQPTs, characterized by critical momenta $k_c$ and critical times $t_c$. Analogous to the Lee-Yang zeros of thermodynamic functions \cite{yang1952, lee1952}, the time variable $t$ in $\mathcal{G}(t)$ are generalized into the complex plane through analytic continuation $t \mapsto z=\tau + it$ \cite{heyl2013, review2016, review2018}, yielding the complex function
\begin{align}
    \cos \left[h_{1}(k)t\right]&=\cos \left[h_{1}(k)(-i)it\right]=\cos \left[ih_{1}(k) it\right]=\cosh \left[h_{1}(k)it\right]\mapsto \cosh \left[h_{1}(k)z\right], \nonumber \\
    i\sin \left[h_{1}(k)t\right]&=i\sin \left[h_{1}(k)(-i)it\right]=-i\sin \left[ih_{1}(k) it\right]=\sinh \left[h_{1}(k)it\right]\mapsto \sinh \left[h_{1}(k)z\right], \nonumber \\
    \mathcal{G}_{k}(z)&=\cosh \left[h_{1}(k)z\right]+ \sinh \left[h_{1}(k)z\right] n_{k}\hat{h}_{0} \cdot \hat{h}_{1}.
\end{align}
Consequently, the zeros of $\mathcal{G}_{k}(z)$ in the complex plane, known as Fisher zeros, are calculated as
\begin{align}
    z_{n}&=\frac{1}{h_{1}(k)}\tanh^{-1}\left(-\frac{1}{n_{k}\hat{h}_{0} \cdot \hat{h}_{1}}\right) \nonumber\\
    &=\frac{1}{2h_{1}(k)}\ln\left(\frac{1-\frac{1}{n_{k}\hat{h}_{0} \cdot \hat{h}_{1}}}{1+\frac{1}{n_{k}\hat{h}_{0} \cdot \hat{h}_{1}}}\right) \nonumber\\
    &=\frac{1}{2h_{1}(k)}\ln\left(\frac{n_{k}\hat{h}_{0} \cdot \hat{h}_{1}-1}{n_{k}\hat{h}_{0} \cdot \hat{h}_{1}+1}\right) \nonumber\\
    &=\frac{1}{2h_{1}(k)}\ln\left(e^{i(2n+1)\pi}\frac{1-n_{k}\hat{h}_{0} \cdot \hat{h}_{1}}{1+n_{k}\hat{h}_{0} \cdot \hat{h}_{1}}\right) \nonumber\\
    &=\frac{i(2n+1)\pi}{2h_{1}(k)}+\frac{1}{2h_{1}(k)}\ln\left(\frac{1-n_{k}\hat{h}_{0} \cdot \hat{h}_{1}}{1+n_{k}\hat{h}_{0} \cdot \hat{h}_{1}}\right) \nonumber\\
    &=\frac{i\pi(2n+1)}{2 h_{1}(k)}-\frac{1}{h_{1}(k)}\tanh^{-1}\left(n_{k}\hat{h}_{0} \cdot \hat{h}_{1}\right), \quad n \in \mathbb{Z},
\end{align}
Here, we have used the relation $\tanh^{-1}(x)=\ln\left[(1+x)/(1-x)\right]/2$ (the principal value).
By introducing the notation $\mathcal{Q}_{k}=\tanh^{-1}\left(n_{k}\hat{h}_{0} \cdot \hat{h}_{1}\right)$, $a=\pi(n+1/2)$, and decomposing 
$h_{1}(k)$ and $\mathcal{Q}_{k}$ into their real and imaginary parts as $h_{1r}=\mathrm{Re}\left[h_{1}(k)\right],h_{1i}=\mathrm{Im}\left[h_{1}(k)\right],\mathcal{Q}_{kr}=\mathrm{Re}\left[\mathcal{Q}_{k}\right], \mathcal{Q}_{ki}=\mathrm{Im}\left[\mathcal{Q}_{k}\right]$, the Fisher zeros can be simplified to
\begin{align}
    z_{n}=\frac{1}{|h_{1}(k)^{2}|}\left[(a h_{1i}-h_{1r}\mathcal{Q}_{kr}-h_{1i}\mathcal{Q}_{ki})+i(a h_{1r}-h_{1r}\mathcal{Q}_{ki}+h_{1i}\mathcal{Q}_{kr})\right].
\end{align}
The physical critical points of DQPTs are determined by the Fisher zeros lying on the imaginary axis, which simultaneously identify the critical momenta 
$k_c$ through the condition
\begin{align}
    a h_{1i}-h_{1r}\mathcal{Q}_{kr}-h_{1i}\mathcal{Q}_{ki}=0.
\end{align}
This condition implies that the two vectors $\vec{v}_{s}=(\mathcal{Q}_{kr},\mathcal{Q}_{ki}-a)$ and $\vec{v}_{h}=(h_{1r},h_{1i})$ in the 2D plane are orthogonal, i.e., $\vec{v}_{s}\cdot \vec{v}_{h}=0$.

\subsection{Non-biorthogonal formulation}
The Loschmidt amplitude admits an alternative definition via non-biorthogonal density matrices (Eq. (\ref{suppeqnbdm})). Its pure-state formulation
\begin{align}
    \mathcal{G}(t)=\operatorname{Tr}\left[\rho(0)U(t)\right]=\operatorname{Tr}\left[\ket{\Psi_{0}}\bra{\Psi_{0}}e^{-i\hat{H}_{1}t}\right],
\end{align}
actually provides a more direct physical interpretation. Provide the initial pure state $\ket{\Psi_{0}}=\bigotimes_{(n,k)\in occ} \ket{u_{n}^{0}(k)}$, the Loschmidt amplitude is calculated by
\begin{align}
    \mathcal{G}(t)&=\prod_{(n,k) \in occ}\prod_{(n',k')}\prod_{k_{0}\in BZ}\langle\!\langle u_{n'}^{0}(k')\ket{u_{n}^{0}(k)}\bra{u_{n}^{0}(k)}e^{-i\hat{c}^{\dagger}_{k_0}H_{1}(k_0)\hat{c}_{k_0}t}\ket{u_{n'}^{0}(k')} \nonumber\\
    &=\prod_{(n,k)\in occ}\braket{u_{n}^{0}(k)|e^{-iH_{1}(k)t}|u_{n}^{0}(k)}\nonumber\\
    &=\prod_{(n,k)\in occ}\braket{u_{n}^{0}(k)|u_{n}^{0}(k,t)}.
\end{align}
The above expression is conventionally normalized by the state norms, yielding
\begin{align}
    \mathcal{G}(t)&=\prod_{(n,k)\in occ}\frac{\braket{u_{n}^{0}(k)|u_{n}^{0}(k,t)}}{\left|\ket{u_{n}^{0}(k)}\right|\left|\ket{u_{n}^{0}(k,t)}\right|}\nonumber\\
    &=\prod_{(n,k)\in occ}\frac{\braket{u_{n}^{0}(k)|u_{n}^{0}(k,t)}}{\sqrt{\braket{u_{n}^{0}(k)|u_{n}^{0}(k)}}\sqrt{\braket{u_{n}^{0}(k,t)|u_{n}^{0}(k,t)}}},
\end{align}
where the corresponding Loschmidt echo, defined as 
\begin{align}
    \mathcal{L}(t)=\left|\mathcal{G}(t)\right|^{2}=\prod_{(n,k)\in occ}\frac{\left|\braket{u_{n}^{0}(k)|u_{n}^{0}(k,t)}\right|^{2}}{\braket{u_{n}^{0}(k)|u_{n}^{0}(k)}\braket{u_{n}^{0}(k,t)|u_{n}^{0}(k,t)}},
\end{align}
is commonly referred to as the self-norm Loschmidt echo \cite{lu2025dyn}. Such Loschmidt echo carries a clear physical interpretation: it quantifies the overlap between the time-evolved postquench state and the initial eigenstate, characterizing the return probability following a quantum quench.

In the general mixed-state scenario, the initial density matrix resembles the biorthogonal form (Eq. (\ref{suppeqbdensity})), with the key distinction lying in whether a biorthogonal basis is employed for its representation,
\begin{align}
   \label{suppeqnbdensity}
    \rho(0)&=\sum_{\alpha}p_{\alpha}\left(\bigotimes_{k,n}\ket{u_{n}^{0}(k)}\right)\left(\bigotimes_{k,n}\bra{u_{n}^{0}(k)}\right) \nonumber\\
    &=\sum_{\left\{\tilde{k},\tilde{n}\right\}}\bigotimes_{j}p_{k_j}^{n_j}\ket{u_{n_j}^{0}(k_j)}\bra{u_{n_j}^{0}(k_j)}\nonumber\\
    &=\sum_{\left\{\tilde{k}\right\}}\sum_{\left\{\tilde{n}\right\}}\bigotimes_{j}p_{k_j}^{n_j}\ket{u_{n_j}^{0}(k_j)}\bra{u_{n_j}^{0}(k_j)}\nonumber\\
    &=\sum_{\left\{\tilde{k}\right\}}\bigotimes_{j}\sum_{n}p_{k_j}^{n}\ket{u_{n}^{0}(k_j)}\bra{u_{n}^{0}(k_j)}\nonumber\\
     &=\sum_{\left\{\tilde{k}\right\}}\bigotimes_{j}\rho_{k_j}(0),
\end{align}
where $p_\alpha=\prod_{j}p_{k_j}^{n_j}$ and $\rho_{k_j}(0)\equiv\sum_{n}p_{k_j}^{n}\ket{u_{n}^{0}(k_j)}\bra{u_{n}^{0}(k_j)}$. Assuming that all band indices $n$ are occupied for fermionic systems, the Loschmidt amplitude for a particular non-repeating momentum occupation is given by (up to a normalization factor)
\begin{align}
    \mathcal{G}(t)&=\mathrm{Tr}\left[\rho(0)U(t)\right]\nonumber\\
    &=\mathrm{Tr}\left[\bigotimes_{k}\rho_{k}(0)\prod_{k_0}e^{-i \hat{c}_{k_{0}}^{\dagger}H_{1}(k_{0})\hat{c}_{k_{0}}t}\right]\nonumber\\
    &=\prod_{k}\mathcal{G}_{k}(t),
\end{align}
where 
\begin{align}
    \mathcal{G}_{k}(t)&=\mathrm{Tr}\left[\rho_{k}(0)U_{k}(t)\right]\nonumber\\
    &=\sum_{k',n'}\langle\!\langle u_{n'}(k')|\left[\sum_{n}p_{k}^{n}\ket{u_{n}^{0}(k)}\bra{u_{n}^{0}(k)}e^{-i \hat{c}_{k_{0}}^{\dagger}H_{1}(k_{0})\hat{c}_{k_{0}}t}\right]\ket{u_{n'}(k')}\nonumber\\
    &=\sum_{n}p_{k}^{n}\bra{u_{n}^{0}(k)}e^{-i H_{1}(k)t} \ket{u_{n}^{0}(k)}\nonumber\\
    &=\sum_{n}p_{k}^{n}\braket{u_{n}^{0}(k)| u_{n}^{0}(k,t)}, \\
    U_{k}(t)&=e^{-i H_{1}(k)t}.
\end{align} 
The corresponding mixed-state Loschmidt echo is given by
\begin{align}
    \mathcal{L}(t)=\left|\mathcal{G}(t)\right|^{2}=\prod_{k}\sum_{n,n'}p_{k}^{n*}p_{k}^{n'}\braket{u_{n}^{0}(k,t)|u_{n}^{0}(k)}\braket{u_{n'}^{0}(k)|u_{n'}^{0}(k,t)}.
\end{align}  
Likewise, the overall normalization factor of the Loschmidt amplitude or echo can be safely omitted.

As a further illustration, we consider again the two-band model $\mathcal{H}_{0}(k)=\vec{h}_{0}(k)\cdot \vec{\sigma}$. However, in the non-Hermitian context, a non-biorthogonal mixed state such as the Gibbs state---with analytically tractable participation probabilities $p_{k}^{\pm}$---is generally not applicable. This is because the natural extension of statistical mechanics to non-Hermitian systems adopts the biorthogonal rather than the non-biorthogonal framework. The Loschmidt amplitude formally corresponds to 
\begin{align}
    \mathcal{G}_{k}(t)&=p_{k}^{+}\braket{\tilde{u}_{+}^{0}(k)|e^{-i\mathcal{H}_{1}(k)t}|\tilde{u}_{+}^{0}(k)}+p_{k}^{-}\braket{\tilde{u}_{-}^{0}(k)|e^{-i\mathcal{H}_{1}(k)t}|\tilde{u}_{-}^{0}(k)} \nonumber\\
    &=p_{k}^{+}\braket{\tilde{u}_{+}^{0}(k)|\cos\left[h_{1}(k)t\right]\mathbbm{1}-i\sin\left[h_{1}(k)t\right]\hat{\mathcal{H}}_{1}(k)|\tilde{u}_{+}^{0}(k)}+p_{k}^{-}\braket{\tilde{u}_{-}^{0}(k)|\cos\left[h_{1}(k)t\right]\mathbbm{1}-i\sin\left[h_{1}(k)t\right]\hat{\mathcal{H}}_{1}(k)|\tilde{u}_{-}^{0}(k)}\nonumber\\
    &=p_{k}^{+}\left(\cos\left[h_{1}(k)t\right]-i\sin\left[h_{1}(k)t\right]\braket{\tilde{u}_{+}^{0}(k)|\hat{\mathcal{H}}_{1}(k)|\tilde{u}_{+}^{0}(k)}\right)+p_{k}^{-}\left(\cos\left[h_{1}(k)t\right]-i\sin\left[h_{1}(k)t\right]\braket{\tilde{u}_{-}^{0}(k)|\hat{\mathcal{H}}_{1}(k)|\tilde{u}_{-}^{0}(k)}\right) \nonumber\\
    &=\cos\left[h_{1}(k)t\right](p_{k}^{+}+p_{k}^{-})-i\sin\left[h_{1}(k)t\right]\left(p_{k}^{+}\braket{\tilde{u}_{+}^{0}(k)|\hat{\mathcal{H}}_{1}(k)|\tilde{u}_{+}^{0}(k)}+p_{k}^{-}\braket{\tilde{u}_{-}^{0}(k)|\hat{\mathcal{H}}_{1}(k)|\tilde{u}_{-}^{0}(k)}\right)\nonumber\\
    &=(p_{k}^{+}+p_{k}^{-})\left\{\cos\left[h_{1}(k)t\right]-i\sin\left[h_{1}(k)t\right]\left(\tilde{p}_{k}^{+}\braket{\tilde{u}_{+}^{0}(k)|\hat{\mathcal{H}}_{1}(k)|\tilde{u}_{+}^{0}(k)}+\tilde{p}_{k}^{-}\braket{\tilde{u}_{-}^{0}(k)|\hat{\mathcal{H}}_{1}(k)|\tilde{u}_{-}^{0}(k)}\right)\right\}\nonumber\\
    &=(p_{k}^{+}+p_{k}^{-})\left(\cos\left[h_{1}(k)t\right]-i\sin\left[h_{1}(k)t\right]\braket{P}_{0}\right),
\end{align}
where we have denoted $\tilde{p}_{k}^{\pm}=p_{k}^{\pm}/(p_{k}^{+}+p_{k}^{-})$ and $\braket{P}_{0}=\tilde{p}_{k}^{+}\braket{\tilde{u}_{+}^{0}(k)|\hat{\mathcal{H}}_{1}(k)|\tilde{u}_{+}^{0}(k)}+\tilde{p}_{k}^{-}\braket{\tilde{u}_{-}^{0}(k)|\hat{\mathcal{H}}_{1}(k)|\tilde{u}_{-}^{0}(k)}$.
Here, $\ket{\tilde{u}_{\pm}^{0}(k)}$ are the self-norm right eigenstates of $\mathcal{H}_{0}(k)$, satisfying $\braket{\tilde{u}_{\pm}^{0}(k)|\tilde{u}_{\pm}^{0}(k)}=1$, which is inconsistent with the biorthogonal relations. Following the derivation in the biorthogonal formulation, we can obtain the non-biorthogonal Fisher zeros 
\begin{align}
    z'_{n}=\frac{i\pi(2n+1)}{2 h_{1}(k)}+\frac{1}{h_{1}(k)}\tanh^{-1}\braket{P}_{0}, \quad n \in \mathbb{Z}.
\end{align}
Again, by introducing the notation $\mathcal{Q}'_{k}=\tanh^{-1}\braket{P}_{0}$, $a=\pi(n+1/2)$, and decomposing 
$h_{1}(k)$ and $\mathcal{Q}'_{k}$ into their real and imaginary parts as $h_{1r}=\mathrm{Re}\left[h_{1}(k)\right],h_{1i}=\mathrm{Im}\left[h_{1}(k)\right],\mathcal{Q}'_{kr}=\mathrm{Re}\left[\mathcal{Q}'_{k}\right], \mathcal{Q}'_{ki}=\mathrm{Im}\left[\mathcal{Q}'_{k}\right]$, the Fisher zeros can be simplified to
\begin{align}
    z'_{n}=\frac{1}{|h_{1}(k)^{2}|}\left[(a h_{1i}+h_{1r}\mathcal{Q}'_{kr}+h_{1i}\mathcal{Q}'_{ki})+i(a h_{1r}+h_{1r}\mathcal{Q}'_{ki}-h_{1i}\mathcal{Q}'_{kr})\right].
\end{align}
The physical critical points of DQPTs are determined by the Fisher zeros lying on the imaginary axis, which simultaneously identify the critical momenta 
$k_c$ through the condition
\begin{align}
    a h_{1i}+h_{1r}\mathcal{Q}'_{kr}+h_{1i}\mathcal{Q}'_{ki}=0.
\end{align}
This condition implies that the two vectors $\vec{v}'_{s}=(\mathcal{Q}'_{kr},\mathcal{Q}'_{ki}+a)$ and $\vec{v}_{h}=(h_{1r},h_{1i})$ in the 2D plane are orthogonal, i.e., $\vec{v}'_{s}\cdot \vec{v}_{h}=0$.

\section{Non-Hermitian winding number}
\label{nhwinding}
Consider a 1D non-Hermitian sublattice (chiral) symmetric two-band Hamiltonian 
\begin{align}
    \mathcal{H}_{s}(k)=d_x\sigma_x+d_y \sigma_y = \left(\begin{matrix}
       0 & d_x-i d_y \\
        d_x + id_y & 0
    \end{matrix}\right)=\left(\begin{matrix}
       0 & A+iB \\
       C+iD & 0
    \end{matrix}\right),
\end{align}
where
\begin{align}
    A&=d_{xr}+d_{yi},\\
    B&=d_{xi}-d_{yr},\\
    C&=d_{xr}-d_{yi},\\
    D&=d_{xi}+d_{yr}.\\
\end{align}
The energies and corresponding biorthogonal eigenstates are
\begin{align}
    &E_{\pm}=\pm\sqrt{(A+iB)(C+iD)}\equiv \pm |d|,\\
    &\ket{u_{\pm}(k)}=\frac{1}{\sqrt{2}}\left(\begin{matrix}
       \frac{A+iB}{E_{\pm}}\\
        1
    \end{matrix}\right)=\frac{1}{\sqrt{2}}\left(\begin{matrix}
       \frac{d_{x}-id_{y}}{\pm |d|}\\
        1
    \end{matrix}\right),\\
    &\langle\!\langle u_{\pm}(k)|=\frac{1}{\sqrt{2}}\left(\begin{matrix}
       \frac{C+iD}{E_{\pm}} & 1
    \end{matrix}\right)=\frac{1}{\sqrt{2}}\left(\begin{matrix}
       \frac{d_{x}+id_{y}}{\pm|d|} & 1
    \end{matrix}\right).
\end{align}

The non-Hermitian winding number is given by \cite{yin2018geometric}
\begin{align}
    \nu=\frac{1}{2\pi}\oint_{BZ}dk \frac{d_x \partial_k d_y -d_y \partial_k d_x}{d_x^2+d_y^2},
\end{align}
and the corresponding Zak phase is derived from the occupied band
\begin{align}
    \gamma=\int_{0}^{2\pi}dk\frac{\langle\!\langle u_{-}(k)|i\partial_{k}\ket{u_{-}(k)}}{\langle\!\langle u_{-}(k)\ket{u_{-}(k)}}=\nu \pi.
\end{align}
Defining 
\begin{align}
    Q_{1}(k)&=d_x-i d_y=d_{xr}+d_{yi}+i(d_{xi}-d_{yr}), \\
    Q_{2}(k)&=d_x +i d_y=d_{xr}-d_{yi}+i(d_{xi}+d_{yr}),
\end{align}
and two associated winding numbers 
\begin{align}
    \nu_1&=\frac{1}{2\pi i}\oint dk Q_{1}(k)^{-1}\partial_k Q_{1}(k)=\frac{1}{2\pi i}\oint dk \frac{(d_{x}+id_{y})(\partial_{k}d_{x}-i\partial_{k}d_{y})}{d_{x}^{2}+d_{y}^{2}}=\frac{1}{2\pi i}\oint dk \frac{d_{x}\partial_{k}d_{x}+d_{y}\partial_{k}d_{y}+id_{y}\partial_{k}d_{x}-id_{x}\partial_{k}d_{y}}{d_{x}^{2}+d_{y}^{2}},\nonumber \\
    \nu_2&=\frac{1}{2\pi i}\oint dk Q_{2}(k)^{-1}\partial_k Q_{2}(k)=\frac{1}{2\pi i}\oint dk \frac{(d_{x}-id_{y})(\partial_{k}d_{x}+i\partial_{k}d_{y})}{d_{x}^{2}+d_{y}^{2}}=\frac{1}{2\pi i}\oint dk \frac{d_{x}\partial_{k}d_{x}+d_{y}\partial_{k}d_{y}-id_{y}\partial_{k}d_{x}+id_{x}\partial_{k}d_{y}}{d_{x}^{2}+d_{y}^{2}},
\end{align}
we immediately obtain 
\begin{align}
    \nu = \frac{\nu_{2}-\nu_{1}}{2}.
\end{align}
On the other hand, $\nu$ can also be defined as the winding number of the real part of the angle $\phi=\tan^{-1}\frac{d_{y}}{d_{x}}$, satisfying
\begin{align}
    &-i\frac{e^{2i\phi}-1}{e^{2i\phi}+1}=-i\frac{e^{i\phi}-e^{-i\phi}}{e^{i\phi}+e^{-i\phi}}=\frac{\sin \phi}{\cos\phi}=\tan\phi=\frac{d_{y}}{d_{x}} \nonumber\\
    &\Rightarrow e^{2i\phi}=\frac{1+i\tan\phi}{1-i\tan\phi}=\frac{1+i\frac{d_{y}}{d_{x}}}{1-i\frac{d_{y}}{d_{x}}} \nonumber\\
    &\Rightarrow \phi=\frac{1}{2i}\ln \frac{d_{x}+id_{y}}{d_{x}-id_{y}},
\end{align}
thus,
\begin{align}
    \nu&=\frac{1}{2\pi}\oint dk \partial_{k}\phi \nonumber\\
    &=\frac{1}{4\pi i}\oint dk \partial_{k}\ln \frac{d_{x}+id_{y}}{d_{x}-id_{y}} \nonumber\\
    &=\frac{1}{4\pi i}\oint dk \frac{d_{x}-id_{y}}{d_{x}+id_{y}}\partial_{k}\frac{d_{x}+id_{y}}{d_{x}-id_{y}}\nonumber\\
    &=\frac{1}{4\pi i}\oint dk \frac{d_{x}-id_{y}}{d_{x}+id_{y}}\left[\frac{\partial_{k}d_{x}+i\partial_{k}d_{y}}{d_{x}-id_{y}}-\frac{d_{x}+id_{y}}{(d_{x}-id_{y})^{2}}(\partial_{k}d_{x}-i\partial_{k}d_{y})\right]\nonumber\\
    &=\frac{1}{4\pi i}\oint dk \left[\frac{\partial_{k}d_{x}+i\partial_{k}d_{y}}{d_{x}+id_{y}}-\frac{\partial_{k}d_{x}-i\partial_{k}d_{y}}{d_{x}-id_{y}}\right]
    \nonumber\\
    &=\frac{1}{4\pi i}\oint dk \left[\frac{(\partial_{k}d_{x}+i\partial_{k}d_{y})(d_{x}-id_{y})}{d_{x}^{2}+d_{y}^{2}}-\frac{(\partial_{k}d_{x}-i\partial_{k}d_{y})(d_{x}+id_{y})}{d_{x}^{2}+d_{y}^{2}}\right]\nonumber\\
    &=\frac{1}{4\pi i}\oint dk \frac{(d_{x}\partial_{k}d_{x}+d_{y}\partial_{k}d_{y}+id_{x}\partial_{k}d_{y}-id_{y}\partial_{k}d_{x})-(d_{x}\partial_{k}d_{x}+d_{y}\partial_{k}d_{y}-id_{x}\partial_{k}d_{y}+id_{y}\partial_{k}d_{x})}{d_{x}^{2}+d_{y}^{2}}
    \nonumber\\
    &=\frac{1}{2\pi }\oint dk \frac{d_{x}\partial_{k}d_{y}-d_{y}\partial_{k}d_{x}}{d_{x}^{2}+d_{y}^{2}}.
\end{align}
Remarkably, only the real part $\phi_{r}$ of $\phi=\phi_{r}+i\phi_{i}$ contributes to the winding number.
Further, $\nu_{1}$ is the winding number of the angle $\tan^{-1}\frac{d_{xi}-d_{yr}}{d_{xr}+d_{yi}}\equiv \phi_{1}$ and $\nu_{2}$ is the winding number of the angle $\tan^{-1}\frac{d_{xi}+d_{yr}}{d_{xr}-d_{yi}}\equiv \phi_{2}$, we also obtain 
\begin{align}
    \nu=\frac{\nu_2-\nu_1}{2}=\frac{1}{2\pi}\oint dk \frac{\partial_k (\phi_{2}- \phi_{1})}{2},
\end{align}
and thus, $2\phi_{r}=\phi_2-\phi_1+2n \pi, n\in \mathbb{Z}$. Noteworthily, the above non-Hermitian winding number formulas are the natural generalization of that in Hermitian systems. Importantly, the two associated winding numbers $\nu_1$ and $\nu_2$ are both integer-quantized due to the argument integrals of $Q_{1}(k)$ and $Q_{2}(k)$ over the BZ, thereby giving rise to a half-integer quantization of $\nu$. In the Hermitian limit, where $\nu_2 = -\nu_1$, one recovers $\nu = \nu_2$, which is also integer-quantized.

\section{Non-biorthogonal DQPTs with winding topology}
Quenching from a chiral symmetric non-Hermitian Hamiltonian $\mathcal{H}_{0}(k)$ to a Hermitian one $\mathcal{H}_{1}(k)$ corresponding to real $h_{1}(k)$, the critical-$k_c$ condition with ground-state like participation probabilities $p_{k}^{-}=1, p_{k}^{+}=0$ reduces to $\braket{\tilde{u}_{-}^{0}(k)|\hat{\mathcal{H}}_{1}(k)|\tilde{u}_{-}^{0}(k)}=0$. Here, we will denote the parameters and energy of $\mathcal{H}_{0}(k)$ as $A_{0}, B_{0}, C_{0}, D_{0}$ and $E_{\pm}^{0}$, and those with $\mathcal{H}_{1}(k)$ as real-value $d_{x}, d_{y}$ and $\pm |d|$. 
Substituting the eigenstates $\ket{u_{\pm}^{0}(k)}$ into this condition, we obtain 
\begin{align}
    \braket{u_{-}^{0}(k)|u_{-}^{0}(k)}\braket{\tilde{u}_{-}^{0}(k)|\hat{\mathcal{H}}_{1}(k)|\tilde{u}_{-}^{0}(k)}&=\braket{u_{-}^{0}(k)|\hat{\mathcal{H}}_{1}(k)|u_{-}^{0}(k)} \nonumber\\
    &=\frac{1}{2|d|}\left(\begin{matrix}
        \frac{A_{0}-iB_{0}}{E_{-}^{0*}} & 1
    \end{matrix}\right)
    \left(\begin{matrix}
        0 & d_{x}-id_{y} \\
        d_{x}+id_{y} & 0
    \end{matrix}\right)
    \left(\begin{matrix}
        \frac{A_{0}+iB_{0}}{E_{-}^{0}} \\
        1
    \end{matrix}\right) \nonumber\\
    &=\frac{1}{2|d|}\left(\begin{matrix}
        d_{x}+id_{y} & \frac{(d_{x}-id_{y})(A_{0}-iB_{0})}{E_{-}^{0*}}
    \end{matrix}\right)
    \left(\begin{matrix}
        \frac{A_{0}+iB_{0}}{E_{-}^{0}} \\
        1
    \end{matrix}\right) \nonumber\\
    &=\frac{1}{2|d|}\left[\frac{(d_{x}+id_{y})(A_{0}+iB_{0})}{E_{-}^{0}}+\frac{(d_{x}-id_{y})(A_{0}-iB_{0})}{E_{-}^{0*}}\right] \nonumber\\
    &=\frac{1}{4|d|^{2}}\left[2|d|\frac{(d_{x}+id_{y})(A_{0}+iB_{0})}{E_{-}^{0}}+2|d|\frac{(d_{x}-id_{y})(A_{0}-iB_{0})}{E_{-}^{0*}}\right] \nonumber\\
    &=\frac{1}{4|d|^{2}}\Bigg[\left(\frac{(d_{x}-id_{y})(A_{0}-iB_{0})}{E_{-}^{0*}}+|d|\right)\left(\frac{(d_{x}+id_{y})(A_{0}+iB_{0})}{E_{-}^{0}}+|d|\right) \nonumber\\
    &\quad -\left(-\frac{(d_{x}-id_{y})(A_{0}-iB_{0})}{E_{-}^{0*}}+|d|\right)\left(-\frac{(d_{x}+id_{y})(A_{0}+iB_{0})}{E_{-}^{0}}+|d|\right)\Bigg] \nonumber\\
    &=\frac{1}{4|d|^{2}}\left[\left|\frac{(d_{x}+id_{y})(A_{0}+iB_{0})}{E_{-}^{0}}+|d|\right|^{2}-\left|-\frac{(d_{x}+id_{y})(A_{0}+iB_{0})}{E_{-}^{0}}+|d|\right|^{2}\right] \nonumber\\
    &=\frac{1}{4}\left|\frac{(d_{x}+id_{y})(A_{0}+iB_{0})}{|d|E_{-}^{0}}+1\right|^{2}-\frac{1}{4}\left|-\frac{(d_{x}+id_{y})(A_{0}+iB_{0})}{|d|E_{-}^{0}}+1\right|^{2},
\end{align}
which leads to the critical-$k_c$ condition 
\begin{align}
    \label{suppdqptc}
    \left|\frac{(d_{x}+id_{y})(A_{0}+iB_{0})}{|d|E_{-}^{0}}+1\right|^{2}&=\left|-\frac{(d_{x}+id_{y})(A_{0}+iB_{0})}{|d|E_{-}^{0}}+1\right|^{2}, \nonumber\\
    \left|\frac{(\hat{d}_{x}+i\hat{d}_{y})(A_{0}+iB_{0})}{E_{-}^{0}}+1\right|^{2}&=\left|-\frac{(\hat{d}_{x}+i\hat{d}_{y})(A_{0}+iB_{0})}{E_{-}^{0}}+1\right|^{2}, \nonumber\\
    \left|\frac{(\hat{d}_{x}+i\hat{d}_{y})+\frac{E_{-}^{0}}{A_{0}+iB_{0}}}{\frac{E_{-}^{0}}{A_{0}+iB_{0}}}\right|^{2}&=\left|-\frac{(\hat{d}_{x}+i\hat{d}_{y})+\frac{E_{-}^{0}}{A_{0}+iB_{0}}}{\frac{E_{-}^{0}}{A_{0}+iB_{0}}}\right|^{2}, \nonumber\\
    \left|(\hat{d}_{x}+i\hat{d}_{y})+\frac{E_{-}^{0}}{A_{0}+iB_{0}}\right|^{2}&=\left|-(\hat{d}_{x}+i\hat{d}_{y})+\frac{E_{-}^{0}}{A_{0}+iB_{0}}\right|^{2}.
\end{align}
Assuming no EPs with $\frac{E_{-}^{0}}{A_{0}+iB_{0}}=0$, we obtain
\begin{align}
    \frac{E_{-}^{0}}{A_{0}+iB_{0}}&=-\frac{\sqrt{(A_{0}+iB_{0})(C_{0}+iD_{0})}}{A_{0}+iB_{0}} \nonumber\\
    &=-\mathrm{sgn}({s_0})\sqrt{\frac{C_{0}+iD_{0}}{A_{0}+iB_{0}}} \nonumber\\
    &=-\mathrm{sgn}({s_0})\sqrt{\frac{(C_{0}^{2}+D_{0}^{2})^{1/2}e^{i \tan^{-1}\frac{D_{0}}{C_{0}}}}{(A_{0}^{2}+B_{0}^{2})^{1/2}e^{i \tan^{-1}\frac{B_{0}}{A_{0}}}}} \nonumber\\
    \nonumber\\
    &=-\mathrm{sgn}({s_0})\mathcal{R}e^{\frac{i}{2}\left(\tan^{-1}\frac{D_{0}}{C_{0}}-\tan^{-1}\frac{B_{0}}{A_{0}}\right)},
\end{align}
where $\mathrm{sgn}(s_0)$ denotes the possible sign caused by the square root and $\mathcal{R}=\left(\frac{C_{0}^{2}+D_{0}^{2}}{A_{0}^{2}+B_{0}^{2}}\right)^{1/4}$. Note that 
\begin{align}
    \tan^{-1}\frac{B_{0}}{A_{0}}&=\tan^{-1}\frac{d_{xi}^{0}-d_{yr}^{0}}{d_{xr}^{0}+d_{yi}^{0}}=\phi_{1}^{0}, \nonumber\\
    \tan^{-1}\frac{D_{0}}{C_{0}}&=\tan^{-1}\frac{d_{xi}^{0}+d_{yr}^{0}}{d_{xr}^{0}-d_{yi}^{0}}=\phi_{2}^{0},
\end{align}
according to Sec. \ref{nhwinding}, we obtain
\begin{align}
    \nonumber\\
    \frac{E_{-}^{0}}{A_{0}+iB_{0}}=-\mathrm{sgn}({s_0})\mathcal{R}e^{i\frac{\phi_{2}^{0}-\phi_{1}^{0}}{2}}=-\mathrm{sgn}({s_0})\mathcal{R}e^{i\phi^{0}}.
\end{align}
The winding number of the prequench $\mathcal{H}_{0}(k)$ is given by
\begin{align}
    \nu^{0}=\frac{1}{2\pi}\oint dk \partial_{k}\phi^{0}.
\end{align}
Moreover, the winding number of the postquench Hermitian $\mathcal{H}_{1}(k)$ is given by
\begin{align}
    \nu^{1}=\frac{1}{2\pi i}\oint dk \hat{d}^{-1}\partial_{k}\hat{d},
\end{align}
where $\hat{d}=\hat{d}_{x}+i\hat{d}_{y}$.
If we regard $\hat{d}$ and $-\mathrm{sgn}({s_0})\mathcal{R}e^{i\phi^{0}}$ in the complex plane as two vectors in the corresponding 2D plane, the critical-$k_c$ condition in Eq. (\ref{suppdqptc}) implies they are orthogonal to each other. Therefore, when the angle difference $\Delta\phi=\phi^{1}-\phi^{0}$ covers $n\pi/2$ (with odd $n$), the DQPTs must occur. Since the winding numbers of $\mathcal{H}_{0}(k)$ and $\mathcal{H}_{1}(k)$ are quantized as half-integer and integer, respectively, the winding number difference $\Delta \nu=\nu^{1}-\nu^{0}=\mathbb{Z}/2$ is the sufficient condition of the occurrence of DQPTs.

Going beyond the zero-temperature case, we extend our analysis to a generic finite-temperature initial state described by arbitrary participation probabilities $p_{k}^{\pm}$, which can, in general, be complex-valued. The according calculation for the upper band is given by   
\begin{align}
    \braket{u_{+}^{0}(k)|u_{+}^{0}(k)}\braket{\tilde{u}_{+}^{0}(k)|\hat{\mathcal{H}}_{1}(k)|\tilde{u}_{+}^{0}(k)}&=\braket{u_{+}^{0}(k)|\hat{\mathcal{H}}_{1}(k)|u_{+}^{0}(k)} \nonumber\\
    &=\frac{1}{2|d|}\left(\begin{matrix}
        \frac{A_{0}-iB_{0}}{E_{+}^{0*}} & 1
    \end{matrix}\right)
    \left(\begin{matrix}
        0 & d_{x}-id_{y} \\
        d_{x}+id_{y} & 0
    \end{matrix}\right)
    \left(\begin{matrix}
        \frac{A_{0}+iB_{0}}{E_{+}^{0}} \\
        1
    \end{matrix}\right) \nonumber\\
    &=\frac{1}{2|d|}\left(\begin{matrix}
        d_{x}+id_{y} & \frac{(d_{x}-id_{y})(A_{0}-iB_{0})}{E_{+}^{0*}}
    \end{matrix}\right)
    \left(\begin{matrix}
        \frac{A_{0}+iB_{0}}{E_{+}^{0}} \\
        1
    \end{matrix}\right) \nonumber\\
    &=\frac{1}{2|d|}\left[\frac{(d_{x}+id_{y})(A_{0}+iB_{0})}{E_{+}^{0}}+\frac{(d_{x}-id_{y})(A_{0}-iB_{0})}{E_{+}^{0*}}\right] \nonumber\\
    &=\frac{1}{4|d|^{2}}\left[2|d|\frac{(d_{x}+id_{y})(A_{0}+iB_{0})}{E_{+}^{0}}+2|d|\frac{(d_{x}-id_{y})(A_{0}-iB_{0})}{E_{+}^{0*}}\right] \nonumber\\
    &=\frac{1}{4|d|^{2}}\Bigg[\left(\frac{(d_{x}-id_{y})(A_{0}-iB_{0})}{E_{+}^{0*}}+|d|\right)\left(\frac{(d_{x}+id_{y})(A_{0}+iB_{0})}{E_{+}^{0}}+|d|\right) \nonumber\\
    &\quad -\left(-\frac{(d_{x}-id_{y})(A_{0}-iB_{0})}{E_{+}^{0*}}+|d|\right)\left(-\frac{(d_{x}+id_{y})(A_{0}+iB_{0})}{E_{+}^{0}}+|d|\right)\Bigg] \nonumber\\
    &=\frac{1}{4|d|^{2}}\left[\left|\frac{(d_{x}+id_{y})(A_{0}+iB_{0})}{E_{+}^{0}}+|d|\right|^{2}-\left|-\frac{(d_{x}+id_{y})(A_{0}+iB_{0})}{E_{+}^{0}}+|d|\right|^{2}\right] \nonumber\\
    &=\frac{1}{4}\left|\frac{(d_{x}+id_{y})(A_{0}+iB_{0})}{|d|E_{+}^{0}}+1\right|^{2}-\frac{1}{4}\left|-\frac{(d_{x}+id_{y})(A_{0}+iB_{0})}{|d|E_{+}^{0}}+1\right|^{2}.
\end{align}
Assuming no EPs with $\frac{E_{+}^{0}}{A_{0}+iB_{0}}=0$, we also obtain
\begin{align}
    \frac{E_{+}^{0}}{A_{0}+iB_{0}}&=\frac{\sqrt{(A_{0}+iB_{0})(C_{0}+iD_{0})}}{A_{0}+iB_{0}} \nonumber\\
    &=\mathrm{sgn}({s_0})\sqrt{\frac{C_{0}+iD_{0}}{A_{0}+iB_{0}}} \nonumber\\
    &=\mathrm{sgn}({s_0})\sqrt{\frac{(C_{0}^{2}+D_{0}^{2})^{1/2}e^{i \tan^{-1}\frac{D_{0}}{C_{0}}}}{(A_{0}^{2}+B_{0}^{2})^{1/2}e^{i \tan^{-1}\frac{B_{0}}{A_{0}}}}} \nonumber\\
    \nonumber\\
    &=\mathrm{sgn}({s_0})\mathcal{R}e^{\frac{i}{2}\left(\tan^{-1}\frac{D_{0}}{C_{0}}-\tan^{-1}\frac{B_{0}}{A_{0}}\right)} \nonumber\\
    &=\mathrm{sgn}({s_0})\mathcal{R}e^{i\phi^{0}}.
\end{align}
Denote $\mathcal{N}_{\pm}^{0}=\braket{u_{\pm}^{0}(k)|u_{\pm}^{0}(k)}$ and $\mathcal{A}_{\pm}^{0}=\frac{\tilde{p}_{k}^{\pm}}{4\mathcal{N}_{\pm}^{0}}\left|\frac{(A_{0}+iB_{0})}{E_{\pm}^{0}}\right|^{2}$, we can obtain
\begin{align}
    \braket{\tilde{u}_{-}^{0}(k)|\hat{\mathcal{H}}_{1}(k)|\tilde{u}_{-}^{0}(k)}&=\frac{1}{4\mathcal{N}_{-}^{0}}\left[\left|\frac{(d_{x}+id_{y})(A_{0}+iB_{0})}{|d|E_{-}^{0}}+1\right|^{2}-\left|-\frac{(d_{x}+id_{y})(A_{0}+iB_{0})}{|d|E_{-}^{0}}+1\right|^{2}\right] \nonumber\\
    &=\frac{1}{4\mathcal{N}_{-}^{0}}\left|\frac{(A_{0}+iB_{0})}{E_{-}^{0}}\right|^{2}\left[\left|\frac{(d_{x}+id_{y})}{|d|}+\frac{E_{-}^{0}}{(A_{0}+iB_{0})}\right|^{2}-\left|-\frac{(d_{x}+id_{y})}{|d|}+\frac{E_{-}^{0}}{(A_{0}+iB_{0})}\right|^{2}\right]\nonumber\\
    &=\frac{1}{4\mathcal{N}_{-}^{0}}\left|\frac{(A_{0}+iB_{0})}{E_{-}^{0}}\right|^{2}\left[ \left|(\hat{d}_{x}+i\hat{d}_{y})-\mathrm{sgn}({s_0})\mathcal{R}e^{i\phi^{0}}\right|^{2}-\left|-(\hat{d}_{x}+i\hat{d}_{y})-\mathrm{sgn}({s_0})\mathcal{R}e^{i\phi^{0}}\right|^{2}\right],
\end{align}
\begin{align}
    \braket{\tilde{u}_{+}^{0}(k)|\hat{\mathcal{H}}_{1}(k)|\tilde{u}_{+}^{0}(k)}&=\frac{1}{4\mathcal{N}_{+}^{0}}
    \left[\left|\frac{(d_{x}+id_{y})(A_{0}+iB_{0})}{|d|E_{+}^{0}}+1\right|^{2}-\left|-\frac{(d_{x}+id_{y})(A_{0}+iB_{0})}{|d|E_{+}^{0}}+1\right|^{2}\right]\nonumber\\
    &=\frac{1}{4\mathcal{N}_{+}^{0}}\left|\frac{(A_{0}+iB_{0})}{E_{+}^{0}}\right|^{2}\left[\left|\frac{(d_{x}+id_{y})}{|d|}+\frac{E_{+}^{0}}{(A_{0}+iB_{0})}\right|^{2}-\left|-\frac{(d_{x}+id_{y})}{|d|}+\frac{E_{+}^{0}}{(A_{0}+iB_{0})}\right|^{2}\right]\nonumber\\
    &=\frac{1}{4\mathcal{N}_{+}^{0}}\left|\frac{(A_{0}+iB_{0})}{E_{+}^{0}}\right|^{2}\left[ \left|(\hat{d}_{x}+i\hat{d}_{y})+\mathrm{sgn}({s_0})\mathcal{R}e^{i\phi^{0}}\right|^{2}-\left|-(\hat{d}_{x}+i\hat{d}_{y})+\mathrm{sgn}({s_0})\mathcal{R}e^{i\phi^{0}}\right|^{2}\right]\nonumber\\
    &=\frac{1}{4\mathcal{N}_{+}^{0}}\left|\frac{(A_{0}+iB_{0})}{E_{+}^{0}}\right|^{2}\left[-\left|(\hat{d}_{x}+i\hat{d}_{y})-\mathrm{sgn}({s_0})\mathcal{R}e^{i\phi^{0}}\right|^{2}+\left|-(\hat{d}_{x}+i\hat{d}_{y})-\mathrm{sgn}({s_0})\mathcal{R}e^{i\phi^{0}}\right|^{2}\right],
\end{align}
\begin{align}
    \braket{P}_{0}&=\tilde{p}_{k}^{+}\braket{\tilde{u}_{+}^{0}(k)|\hat{\mathcal{H}}_{1}(k)|\tilde{u}_{+}^{0}(k)}+\tilde{p}_{k}^{-}\braket{\tilde{u}_{-}^{0}(k)|\hat{\mathcal{H}}_{1}(k)|\tilde{u}_{-}^{0}(k)}\nonumber\\
    &=\frac{\tilde{p}_{k}^{+}}{4\mathcal{N}_{+}^{0}}\left|\frac{(A_{0}+iB_{0})}{E_{+}^{0}}\right|^{2}\left[-\left|(\hat{d}_{x}+i\hat{d}_{y})-\mathrm{sgn}({s_0})\mathcal{R}e^{i\phi^{0}}\right|^{2}+\left|-(\hat{d}_{x}+i\hat{d}_{y})-\mathrm{sgn}({s_0})\mathcal{R}e^{i\phi^{0}}\right|^{2}\right] \nonumber\\
    &+\frac{\tilde{p}_{k}^{-}}{4\mathcal{N}_{-}^{0}}\left|\frac{(A_{0}+iB_{0})}{E_{-}^{0}}\right|^{2}\left[ \left|(\hat{d}_{x}+i\hat{d}_{y})-\mathrm{sgn}({s_0})\mathcal{R}e^{i\phi^{0}}\right|^{2}-\left|-(\hat{d}_{x}+i\hat{d}_{y})-\mathrm{sgn}({s_0})\mathcal{R}e^{i\phi^{0}}\right|^{2}\right]\nonumber\\
    &=\mathcal{A}_{+}^{0}\left[-\left|(\hat{d}_{x}+i\hat{d}_{y})-\mathrm{sgn}({s_0})\mathcal{R}e^{i\phi^{0}}\right|^{2}+\left|-(\hat{d}_{x}+i\hat{d}_{y})-\mathrm{sgn}({s_0})\mathcal{R}e^{i\phi^{0}}\right|^{2}\right] \nonumber\\
    &+\mathcal{A}_{-}^{0}\left[ \left|(\hat{d}_{x}+i\hat{d}_{y})-\mathrm{sgn}({s_0})\mathcal{R}e^{i\phi^{0}}\right|^{2}-\left|-(\hat{d}_{x}+i\hat{d}_{y})-\mathrm{sgn}({s_0})\mathcal{R}e^{i\phi^{0}}\right|^{2}\right]\nonumber\\
    &=\left(-\mathcal{A}_{+}^{0}+\mathcal{A}_{-}^{0}\right)\left[\left|(\hat{d}_{x}+i\hat{d}_{y})-\mathrm{sgn}({s_0})\mathcal{R}e^{i\phi^{0}}\right|^{2}-\left|-(\hat{d}_{x}+i\hat{d}_{y})-\mathrm{sgn}({s_0})\mathcal{R}e^{i\phi^{0}}\right|^{2}\right].
\end{align}
Under the assumption that $-\mathcal{A}_{+}^{0}+\mathcal{A}_{-}^{0} \neq 0$ is real, $\mathcal{Q}'_{k}=\tanh^{-1}\braket{P}_{0}$ is necessarily real, implying that non-biorthogonal DQPTs possess identical topological characteristics at zero and finite temperatures for quenches from a chiral-symmetric non-Hermitian Hamiltonian to a Hermitian one.

\section{Dissipation-controlled quantum quench and electrical circuit realization of the non-Hermitian Su-Schrieffer-Heeger model}
{\em \textbf{Quantum quench and open quantum systems.---}} In theoretical practice, an effective non-Hermitian quantum Hamiltonian can be derived from the Lindblad master equation of an open quantum system, which governs the time evolution of the density matrix \cite{song2019damping, liu2020helical, xue2022}
\begin{align}
    \label{suppeqlindblad}
    \frac{d\rho}{dt}=-i\left[H, \rho \right]+\sum_{\mu}\left(2L_{\mu}\rho L_{\mu}^{\dagger}-\left\{L_{\mu}^{\dagger}L_{\mu},\rho\right\}\right),
\end{align}
where $H$ is a Hermitian Hamiltonian, and $L_{\mu}$'s are the Lindblad operators describing quantum jumps induced by the coupling to the environment. This equation exhibits a non-unitary, short-time evolution described by
\begin{align}
    \label{suppeqeff}
    \frac{d\rho}{dt}=-i\left(H_{eff}\rho-\rho H_{eff}^{\dagger}\right),
\end{align}
which is governed effectively by the non-Hermitian Hamiltonian $H_{eff}=H-i\sum_{\mu}L_{\mu}^{\dagger}L_{\mu}$ prior to the occurrence of quantum jump $2L_{\mu}\rho L_{\mu}^{\dagger}$ [Fig. \ref{supfig-jump}(a)]. Obviously, Eq. (\ref{suppeqeff}) describes the non-biorthogonal density matrix governed by non-unitary time evolution, which aligns with the no-jump limit of the approximate Lindblad master equation \cite{sim2023quantum, sim2025onservables}.

Specifically, we apply the above Lindblad process to the Hermitian Su-Schrieffer-Heeger (SSH) model by introducing flux-controlled Lindblad operators (dissipation), which governs the intra- and inter-unit cell hoppings $t_1$ and $t_2$ between sublattices A and B, as illustrated in Fig. \ref{supfig-jump}(b). The Lindblad operators are give by $L_{n}=\sqrt{\gamma}(c_{nA}+e^{i\phi}c_{nB})$ with the effective flux $\phi$. The effective non-Hermitian Hamiltonian $H_{eff}$ exhibits the non-reciprocal intra-unit cell hoppings $t_{1}\mp i\gamma e^{\pm i \phi}$ from the sublattice B (A) toward A (B), despite the presence of irrelevant onsite energy terms. Particularly, setting the flux $\phi=\pi/2$ yields the standard non-Hermitian SSH model discussed in the main text. Thus, by controlling the dissipation through the Lindblad operators, i.e., suddenly changing their parameters, quantum quenching in the non-Hermitian SSH model can be realized. Importantly, abruptly turning off the parameter $\gamma$ leads to DQPTs with topological characteristics, quenching from a non-Hermitian to a Hermitian system.

\begin{figure}
    \centering
    \includegraphics[width=0.8\linewidth]{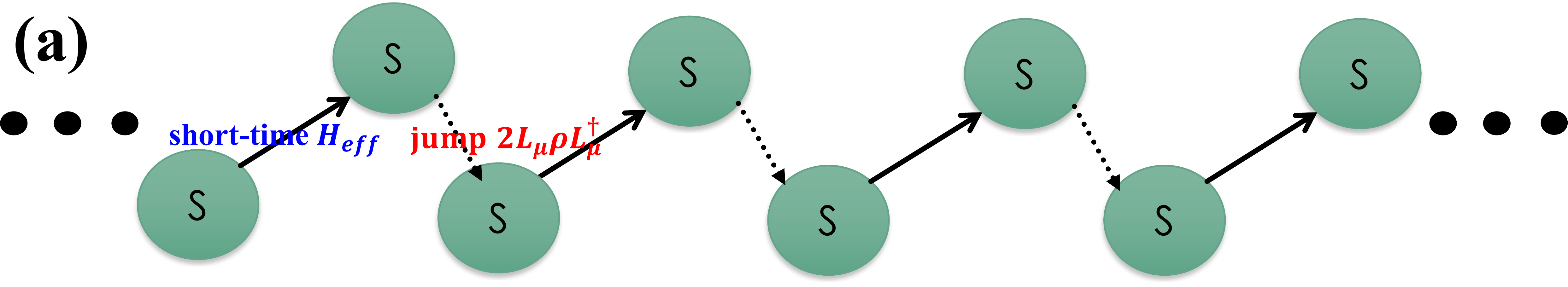}
    \includegraphics[width=0.8\linewidth]{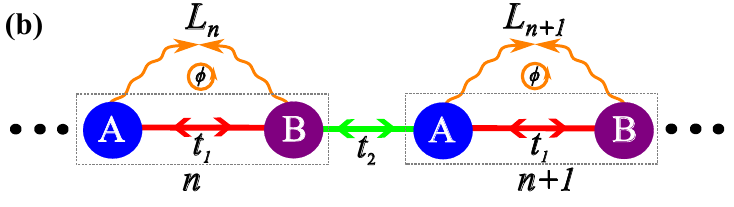}
    \caption{Schematic diagram of dissipation-controlled quantum quench. (a) The effective short-time evolution governed by $H_{eff}$ before the quantum jump $2L_{\mu}\rho L_{\mu}^{\dagger}$, derived from the Lindblad master equation. (b) The flux-controlled Lindblad operators realizing the quantum quench in the non-Hermitian SSH model.}
    \label{supfig-jump}
\end{figure}

{\em \textbf{Circuit realization of non-Hermitian SSH model.---}} While the preceding analysis establishes the theoretical framework for non-Hermitian SSH quenches in open quantum systems, experimental realization requires a different approach. Electrical circuits provide a direct and controllable platform for implementing the non-Hermitian SSH Hamiltonian. Figure \ref{supfig-circuit} shows our circuit design: 
alternating resistors $R_{1,2}$ connect nodes that correspond to sublattices  $A$ and $B$ at site  $n$, with each node grounded through an inductor 
$L$. This configuration naturally encodes the non-Hermitian hopping terms through the impedance converter with current inversion (INIC) \cite{helbig2020, hu2021knots, zhou2025floquet}, denoted by $\pm R_{\gamma}$, indicating that the INIC effectively behaves as a resistor $R_{\gamma}$ ($-R_{\gamma}$) for current flowing from left (right) to right (left), while the inductors set the on-site energy scale. 
The Kirchhoff's law on each node is given by
\begin{align}
    \label{suppeqkir}
    I_{n,A}&=\left(\frac{1}{R_{1}}+\frac{1}{R_{\gamma}}\right)\left(V_{n,A}-V_{n,B}\right)+\frac{1}{R_{2}}\left(V_{n,A}-V_{n-1,B}\right)+\frac{1}{i\omega L}V_{n,A}, \nonumber\\
    I_{n,B}&=\left(\frac{1}{R_{1}}+\frac{1}{-R_{\gamma}}\right)\left(V_{n,B}-V_{n,A}\right)+\frac{1}{R_{2}}\left(V_{n,B}-V_{n+1,A}\right)+\frac{1}{i\omega L}V_{n,B},
\end{align}
where $I_{n,i}$ and $V_{n,i}$, with $i=A,B$, denote the input current and electrical potential at each node, respectively. Under PBCs in the momentum space, the tight-binding circuit model takes the form  
\begin{align}
\mathbf{I}=\mathcal{J}(\omega)\mathbf{V},
\end{align}
where the circuit Laplacian is given by
\begin{align}
    \mathcal{J}(\omega)=\left(
        \begin{matrix}
            \frac{1}{R_{1}}+\frac{1}{R_{2}}+\frac{1}{i\omega L}+\frac{1}{R_{\gamma}} & -\left(\frac{1}{R_{1}}+\frac{1}{R_{\gamma}}+\frac{1}{R_{2}} e^{-ik}\right) \\
            -\left(\frac{1}{R_{1}}-\frac{1}{R_{\gamma}}+\frac{1}{R_{2}} e^{ik}\right) & \frac{1}{R_{1}}+\frac{1}{R_{2}}+\frac{1}{i\omega L}-\frac{1}{R_{\gamma}}
        \end{matrix}
    \right).
\end{align}
It implies 
\begin{align}
 \left(\frac{1}{R_{1}}+\frac{1}{R_{2}}+\frac{1}{i\omega L}\right)\sigma_{0}+\frac{1}{R_{\gamma}} \sigma_{z}-\mathcal{J}(\omega)=\left(\frac{1}{R_{1}}+\frac{1}{R_{2}}\cos k\right)\sigma_{x}+\left(\frac{1}{R_{2}}\sin k +i \frac{1}{R_{\gamma}} \right)\sigma_{y},
\end{align}
where the RHS is exactly the form of the non-Hermitian SSH model. While electrical circuits are capable of implementing non-Hermitian Hamiltonians, the simulation of quantum dynamics and quenches remains a significant challenge, warranting further investigation in future studies.
\begin{figure}
    \centering
    \includegraphics[width=0.8\linewidth]{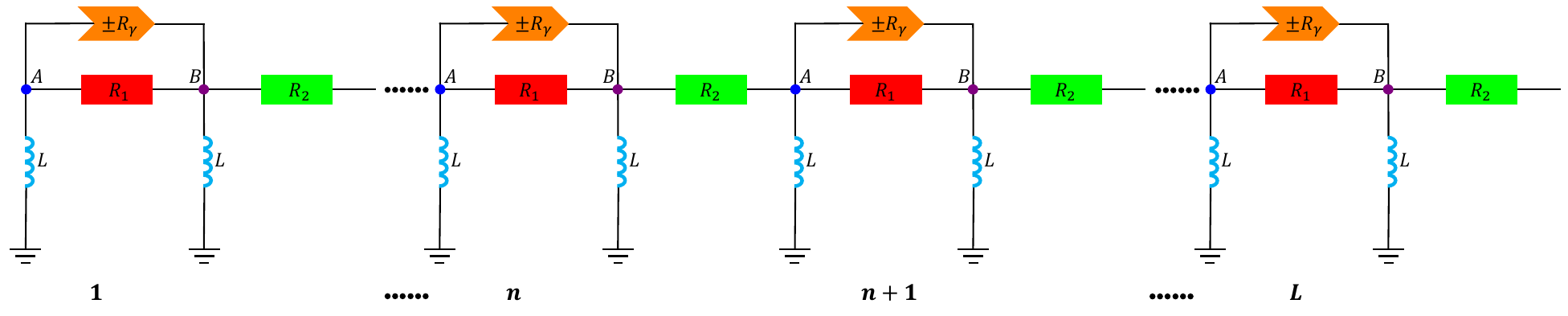}
    \caption{Schematic of the simple circuit design for the non-Hermitian SSH model. Resistors $R_{1,2}$ and grounded inductors $L$ are connected such that the nodes correspond to the sublattices $A$ and $B$ at each site $n$. The non-reciprocal hoppings are realized via INICs.}
    \label{supfig-circuit}
\end{figure}

\end{widetext}

\end{document}